# Filamentary structure of star-forming complexes

**Philip C. Myers**
Harvard-Smithsonian Center for Astrophysics, 60 Garden Street, Cambridge MA 02138 USA

pmyers@cfa.harvard.edu

**Abstract.** The nearest young stellar groups are associated with "hubs" of column density exceeding $10^{22}$ cm$^{-2}$, according to recent observations. These hubs radiate multiple "filaments" of parsec length, having lower column density and fewer stars. Systems with many filaments tend to have parallel filaments with similar spacing. Such "hub-filament structure" is associated with all of the nine young stellar groups within 300 pc, forming low-mass stars. Similar properties are seen in infrared dark clouds forming more massive stars. In a new model, an initial clump in a uniform medium is compressed into a self-gravitating, modulated layer. The outer layer resembles the modulated equilibrium of Schmid-Burgk (1967) with nearly parallel filaments. The filaments converge onto the compressed clump, which collapses to form stars with high efficiency. The initial medium and condensations have densities similar to those in nearby star-forming clouds and clumps. The predicted structures resemble observed hub-filament systems in their size, shape, and column density, and in the appearance of their filaments. These results suggest that hub-filament structure associated with young stellar groups may arise from compression of clumpy gas in molecular clouds.

## 1. Introduction

Many of the nearest star-forming complexes have elongated structures of parsec scale, indicated by images at optical (Barnard 1927, Lynds 1962, Feitzinger & Stüwe 1984, Schneider & Elmegreen 1979, Dobashi et al 2005), infrared (Lada, Alves & Lombardi 2007), and submillimeter wavelengths (Motte, André & Neri 1998), and in rotational lines of CO and its isotopes (Ungerechts & Thaddeus 1987, Mizuno et al 1995, Goldsmith et al 2008). Similar structures are also common in larger and more distant "infrared dark clouds" seen at mid-infrared wavelengths with the *Spitzer Space Telescope* (Devine et al 2004). The prevalence of large-scale filamentary structure suggests that it may persist for a large fraction of a typical cloud lifetime. Such structure may therefore provide clues about the origin and geometry of star-forming regions.

The internal structure of individual filaments has been studied by many authors, but there has been less discussion of the spatial relation of neighboring filaments (Schneider & Elmegreen 1979, Mizuno et al 1995, Lada, Alves & Lada 1999, Hartmann 2002, Goldsmith et al 2008). The relation between central regions and their filamentary extensions has been described as "head-tail" structure, based on CO maps of nearby clouds (Tachihara et al 2002). With deeper observations and finer resolution, it now appears that some "heads" have more than one associated "tail." This paper presents new evidence that multiple parsec-scale filaments tend to branch out from dense star-forming "hubs" in regions forming stellar groups.

The origin of filamentary cloud structure is unclear. Individual filaments can arise from compression of initially uniform gas by converging turbulent flows (Klessen et al 2004, Jappsen et al 2005), and filaments can radiate from hubs following the collision of uniform cylinders of gas (Vazquez-Semadeni et al 2007, Heitsch et al 2008). In a magnetically dominated model, self-gravity pulls gas to the midplane after turbulence dissipates, and ambipolar diffusion allows gravitational instabilities to create filamentary structure in the midplane gas (Nakamura and Li 2008). However, these models do not match the regular spacings and directions of filaments observed in and near hub-filament systems.

This paper applies the basic idea that a self-gravitating layer can develop and preserve filamentary structure, as a consequence of its equilibrium (e.g., Nagai, Inutsuka, & Miyama 1998) or of its collapse (e.g. Burkert & Hartmann 2004). As a specific instance of this idea, this paper presents a new model which matches several observed properties of hub-filament structure. In this model, clumpy molecular cloud gas is compressed into a layer which is vertically self-gravitating and density-modulated, similar to the isothermal equilibrium of Schmid-Burgk (1967). The model reproduces the regular spacing of filamentary arms, their convergence toward hubs, and the column densities of hubs and filaments. However the model is incomplete, and should be considered a starting point for further explorations, especially with numerical simulations.

The paper is organized into six sections. Section 2 describes hub-filament structure in nearby clouds forming low-mass stars, and in more distant clouds forming more massive stars. Section 3 describes models of filamentary structure and introduces the new model. Section 4 gives the choice of input parameters and presents calculated images of column density. Section 5 discusses this model, and Section 6 gives a summary. Expressions for the density and column density of the initial clumpy medium are given in the Appendix.

## 2. Hub-filament structure

This section describes hub-filament structure (HFS) and its properties in star-forming clouds. Here "hub" denotes a central body of low aspect ratio and high column density, and "filament" denotes an associated feature of greater aspect ratio and lower column density.

To show the high incidence of HFS in nearby star-forming regions, subsection 2.1 presents a "complete" survey, describing HFS associated with all nine young stellar groups known within 300 pc. Subsection 2.2 presents seven regions forming more massive stars and also showing HFS. These indicate that HFS is not limited to nearby clouds, and is not limited to regions of low-mass star formation. Subsection 2.3 describes the internal structure of hubs, and Subsection 2.4 gives a summary of HFS properties.

*2.1. The nearest star-forming complexes*

Properties of HFS associated with the nearest young stellar groups are described in the following paragraphs, and are summarized in Table 1.

A useful sample of young stellar groups is defined by the following criteria. A "stellar group" must have at least 20 associated young stellar objects (YSOs) within a projected radius of at most 0.5 pc, corresponding to a minimum surface density of 25 YSOs $pc^{-2}$. A YSO is defined to have spectral class 0, I, transition, or II, according to the system of Lada & Wilking (1984) as revised by Evans et al (2008) and Gutermuth et al (2008a). A stellar group is "young" if the fraction of its YSOs which are Class 0 or I "protostars" is at least 0.1. This restriction excludes stellar groups in IC348, Cha I, and Lupus III which have largely completed their star formation and have dispersed a significant fraction of their original gas. A young stellar group is "nearby" if its distance is less than 300 pc, allowing relatively complete coverage from recent *Spitzer Space Telescope* surveys.

Related definitions of young stellar groups or clusters are given by Lada & Lada (2003), Porras et al (2003), Adams et al (2006), Jørgensen et al (2008), and Gutermuth et al (2008a). The criteria used here would accept as young stellar groups all "tight clusters" identified by Jørgensen et al (2008), but only some of their "tight groups," "loose groups," and "loose clusters."

With these criteria, recent observations indicate nine nearby young stellar groups, detailed in Table 1 and depicted in Figures 1-9.

In Table 1, Column 1 gives the name of the group or its associated hub, whichever is more common. Column 2 gives the number of associated filaments radiating from the hub, according to inspection of the image in Figures 1-9. The number of identified filaments is estimated to be uncertain by ±1. Filaments judged to be too indistinct or confused are not reported. Column 3 gives the number of "neighbor" filaments whose length and direction are similar to those of radiating filaments. Column 4 gives the maximum column density of the hub, based on near-infrared extinction. Column 5 gives the number of YSOs in the group, enclosed by the projected effective radius in Column 6. In L1495 and L1688 there are two neighboring subgroups, so both radii are given. Column 7 gives the fraction of group YSOs which are protostars. Column 8 gives the estimated distance to the region, and Column 9 gives references for the hub and filament images, stellar content, class, and distance. The entries in Table 1 are listed in order of the number of radiating filaments, in Column 2.

Images of the regions in Table 1 are shown in Figures 1-9 of the on-line edition. These images were chosen to reveal extended filamentary structure, from the astronomical literature and from web sites displaying deep, wide-field astrophotography. They are reproduced only in the on-line edition of this paper because their faint structure is best seen in its originally published form, whereas conversion to printed images generally reduces useful information.

Figure 1 shows a deep optical image of the reflection nebula NGC7023, situated in the hub of high extinction in L1174 (Lynds 1962; tvdavisastropics.com). The extinction extends S in L1172 and can be seen for nearly 5 pc, as a single filamentary lane with a small side branch to the E. This hub with a single filament resembles the "head-tail" structure described by Tachihara et al (2002), whereas all of the other examples in this section have hubs which radiate more than one filament. To bring out faint structure in this image and in some of the following images, the darkest and brightest tones were removed.

The Taurus complex is shown in Figure 2, based on observations in the J=1-0 line of $^{13}$CO. The color scheme indicates distinct velocity intervals (Goldsmith et al 2008). In the NW, the hub L1495 radiates the prominent green filament B213 for about 7 pc to the SE and the fainter filaments L1484 and L1486 for about 7 and 4 pc respectively to the NW. Three additional "neighbor filaments" have similar length and orientation to the B213-L1484 system, with projected separation of a few pc, as noted by Goldsmith et al (2008). There are six well-known groupings of young stars in Taurus (Kenyon, Gómez, & Whitney 2008, Gómez et al 1993), but only the densest one, in L1495, meets the present criteria for a young stellar group.

Figure 3 shows an $A_K$ extinction image of Barnard 59, the most prominent part of the Pipe nebula (Lombardi et al 2006). The central region radiates two well-defined curving filaments of length 1 pc to the S and SE, and a fainter curving arm of similar length to the NW. Fainter elongated features are also seen in Figure 3 and in an optical image from the Digitized Sky Survey (skyview.gsfc.nasa.gov), but they are too indistinct to be counted as filaments in Table 1.

Figure 4 shows the group in Serpens South (Gutermuth et al 2008b) in an image made from IRAC bands 1 and 2. The young cluster radiates two thin filaments to the S and SE, and a broader feature to the NW. This NW feature appears to join a second hub-filament system to the N. This second system has no stars in its hub, and its filaments extend further N for about 0.5 pc to a another concentration of young stars, with fewer members than the Serpens South group.

Figure 5 shows 1.1 mm emission from Serpens Cluster A (Enoch et al 2007). The image shows a bright elongated hub with filamentary extensions to the NW and S, and with two faint filaments to the NE. Alternatively the extended emission to the N and to the S could be interpreted as part of an elongated hub radiating two filaments to the NE. However the greatly increased intensity of the central emission supports its identification as a distinct hub, with four radiating filaments. The faint filaments radiating to the NE are also seen in 0.85 mm emission and in CO 2-1 emission, where they have greater extension than in Figure 5 (Davis et al 1999).

Figure 6 shows the IC348 region in the E part of the Perseus complex, in contours of integrated intensity of the J=1-0 line of $^{13}$CO (Ridge et al 2006), with associated YSOs based on *Spitzer Space Telescope* observations (Muench et al 2007). The dense gas emission shows a hub with four radiating filaments, each 0.5-1 pc long. In the higher-resolution 850 $\mu$m image of Kirk, Johnstone and Di Francesco (2006), a faint neighbor filament about 3 pc long lies about 2 pc to the SE. In Figure 6 the concentration of young YSOs in the SW is centered on the hub seen in the $^{13}$CO line, similar to the preceding examples of HFS.

Figure 6 also shows that the main concentration of YSOs in IC348 consists mostly of Class II sources projected on very low levels of $^{13}$CO emission. The protostar fraction in IC348 is 2/56=0.04 (Gutermuth et al 2008a), too low to qualify as a young stellar group in Table 1. The contrast between IC348 and IC348-SW suggests that HFS is more often associated with young groups whose dense gas has not yet dispersed.

The NGC 1333 region in the W part of the Perseus complex is shown in a deep optical image (nightskyphotography.com) in Figure 7. The embedded cluster is seen in an elongated central region, having 5 filamentary extensions each about 1 pc long. NGC 1333 is situated in a lower-density filament extending about 5 pc NE to SW. In 0.85 mm emission low-density filaments of similar length lie to both E and W (Hatchell et al 2005). However their long axis directions differ too much from that of NGC 1333 to be counted in Table 1 as parallel "neighbor filaments."

Figure 8 shows a deep optical image (astromodelismo.es) of the Ophiuchus complex. It shows L1688 as an opaque elongated hub, which radiates four filamentary lanes containing L1712, L1755 and L1757, to the NE. The two northernmost lanes are seen more clearly here than in the star-count images of Dobashi et al (2005) or the $^{13}$CO map of Loren (1989), probably because of finer resolution. However Figure 8 renders the close neighbors L1688 and L1689 as part of the same hub, probably due to high extinction, whereas they appear more separate in other images. The four NE lanes are nearly parallel, and extend for 15-25 pc in the plane of the sky. A neighbor filament to the NW includes L1752, elongated roughly parallel to these lanes but distinct from L1688. Two faint parallel curving lanes, including L1672, extend S from L1688 for 8-10 pc. L1688 has the longest arms, the greatest peak column density, the most embedded YSOs, and has the second greatest number of arms among the nine regions in Table 1.

A deep optical image (starryscapes.com) of the Corona Australis cloud is shown in Figure 9. There, the opaque elongated hub FSD 447 (Feitzinger & Stüwe 1984) appears to radiate five nearly parallel lanes with similar spacing to the NE, and three to the SW. At the NW end of the image, bright nebulosity from the embedded Coronet cluster lies between the opaque hub and the filament elongated NE-SW to the hub. However, maps of integrated intensity of $C^{18}O$ 1-0 indicates that the hub connects smoothly to this filament (Harju et al 1993, Yonekura et al 1999). The third of the five filaments extending to the NE is faint, but is also seen in image UKS37 at aao.gov.au/images. To the S of the hub lies a long translucent filament, which may be connected to the hub, or may be a disconnected "neighbor filament" seen in projection.

The Cr A hub-filament structure resembles that of the Oph cloud in its hub aspect ratio and number of filaments, but differs in having lower peak surface density, shorter filaments, and fewer associated young stars.

Table 1 and Figures 1-9 describe the features of HFS, and indicate that these properties are associated with all nine of the known young stellar groups within 300 pc. They show that in the median case, a nearby young stellar group having ~30 YSOs within a few 0.1 pc is associated with a hub of peak column density $\sim 2 \times 10^{22}$ cm$^{-2}$. A typical hub radiates several filaments which can be followed for several pc down to a minimum discernable column density $\sim 1 \times 10^{21}$ cm$^{-2}$. In three cases, at least one neighboring filament is also seen, having length and direction similar to those of the main HFS features. Groups having more members, such as NGC 1333 in Perseus and L1688 in Ophiuchus, tend to have higher peak column density and more filamentary arms.

The incidence of HFS found in nearby young stellar groups depends on the criteria used to define young stellar groups. If the definition were changed to include older or sparser groups, there would be more groups but their fraction with HFS would decrease. If older groups such as IC348, Lupus III, and Cha I were included the fraction would decline slightly, from 9/9 to 10/12. However if the minimum number of members were decreased from the current value of 20 to five, as used by Jørgensen et al (2008), the fraction would fall to a much lower value, since most such small stellar groups have no noticeable indication of associated HFS. Evidently HFS requires a sufficiently large concentration of gas and young stars to be an associated property.

*2.2. HFS in clouds forming more massive stars*

The nearby stellar groups detailed in Table 1 and Figures 1-9 are regions of low-mass star formation, with no significant H II regions. This section shows that HFS is also seen in more distant regions forming both massive and low-mass stars. These more distant regions appear to have more massive stars, greater column densities, and more radiating filaments. As in Section 2.1, identification and counting of filaments is uncertain, and filaments are not counted if they are judged to be too faint or confused. The incidence of HFS in these more massive star-forming regions is not known, since in most cases, HFS can be seen only in the nearest clouds viewed against the bright infrared background of the galactic bulge.

As with Figures 1-9, Figures 10-15 are shown only in the on-line version of this paper.

Figure 10 shows dust emission 0.85 mm wavelength from the "integral filament" associated with Orion Molecular Cloud 1 (OMC-1, Johnstone & Bally 1999). In this image, a bright elongated hub extends NS for about 0.3 pc and radiates four filaments to the W, three to the N, and an uncertain number to the E and S, including the Orion "bar" in the S. Johnstone & Bally (1999) state that OMC-1 radiates "at least a dozen dusty filaments." The four most distinct filaments to the W have approximately equal spacing and are nearly parallel where they join the hub. Within 0.5 pc radius there are approximately 400 young stellar objects (Hillenbrand & Hartmann 1998) and the peak gas column density is about $10^{23}$ cm$^{-2}$ (Bally et al 1987). The gas column density and number of associated stars are much greater than in the nearby young stellar groups of Table 1 and Figures 1-9.

Figures 11-15 show six infrared dark clouds (IRDCs) having HFS, from the GLIMPSE/MIPSGAL survey using the IRAC and MIPS cameras on the *Spitzer Space Telescope* (Benjamin et al 2003, Carey et al 2005, alienearths.org/glimpse). The IRDCs shown here are typically 3 kpc distant, based on velocities of their molecular lines and a galactic rotation model (Jackson et al 2008a). In Figures 11-15, their structure is seen in absorption in the IRAC bands against the mid-infrared background light of the inner galaxy. Their peak column densities are $\sim 10^{23}$ cm$^{-2}$, comparable to those in Orion (Jackson et al 2008b). The images are presented in order of increasing number of filaments, as in Figures 1-9.

In the three-color images of Figures 11-15, a small circle with primarily red color indicates pointlike emission at 24 $\mu$m brighter than at 8 $\mu$m and 3 $\mu$m. This color excess is believed to arise from dust grains in a protostellar envelope, which absorb radiation from the central source and reradiate at longer wavelengths (Allen et al 2004). With the IRAC and MIPS sensitivities at this distance, a significant protostar detection indicates luminosity $\sim 10^4$ L$_O$ as in G034.4+0.2, an infrared dark cloud harboring three such sources. These sources are considered candidate massive protostars, which have not yet formed detectable H II regions (Rathborne et al 2005). It is expected that each such massive protostar will be associated with a cluster of lower-mass stars, based on the strong association of OB stars with clusters (de Wit et al 2005). Thus the IRDCs in Figures 11-15 appear to resemble OMC-1 in Figure 10, as regions of very high column density forming massive stars and young clusters.

Figures 11-13 show four IRDCs with candidate massive protostars and hub-filament structure, each showing at least four filaments radiating from a central hub. These resemble the lower-

mass systems in Figures 3-6, where star-forming hubs with relatively little elongation radiate filaments with similar spacing but varying directions. Figures 14 and 15 show IRDCs with more elongated hubs and at least six filaments. Their elongated hubs radiate numerous filaments with similar spacing, and with similar direction across the long axis of the hub. These properties are similar to those in Ophiuchus (Figure 8), Corona Australis (Figure 9), and Orion (Figure 10).

*2.3. Hub structure*

The foregoing examples indicate that hubs have high surface density of gas and stars, and radiate external filaments. Describing the internal structure of hubs is more difficult, since it requires observations which are not saturated due to high optical depth, and which are not confused by nebulosity. Even then, dispersal of dense gas by stellar winds and radiation limits what can be deduced from observations. The following discussion is limited to well-studied hubs in Ophiuchus and NGC1333.

In Ophiuchus, the elongated hub extending SE-NW in Figure 8 breaks up into two distinct dense regions, L1688 and L1689, when viewed in images of $^{13}$CO J=1-0 emission (Loren 1989) or star-count extinction (Dobashi et al 2005). In turn each of these appears as denser, elongated features whose long axes have the same SE-NW direction, when viewed with higher-resolution tracers sensitive to higher density. A map of L1688 in $C^{18}O$ 1-0 shows an elongated structure with two similarly elongated peaks (Wilking, Gagné & Allen 2008) while a map of L1689 in 0.85 mm dust continuum emission shows one prominent filament and three fainter ones, all approximately parallel to each other and to the long axis in L1688 (Nutter, Ward-Thompson & André 2006). These four filaments have mean spacing 0.5 pc, finer than the mean spacing 1.3 pc of the four extended filaments in Figure 8. Thus the elongated hub seen in the optical image resolves into parallel filaments which are denser and have finer spacing than those which radiate outward.

In NGC1333, dust emission images at 0.85 mm (Gutermuth et al 2008c) and 1.2 mm (Enoch et al 2006) each show large-scale radiating filaments as in the optical image in Figure 7. In the 0.85 mm image, the hub resolves into four bright peaks, two lying within nearly parallel filaments having length and separation ~0.4 pc. Together these features form a segmented ellipse with major axis ~0.6 pc. Each peak is brighter than the outermost filament contour by a factor 10-30. As in Ophiuchus, the hub resolves into lumpy parallel filaments of higher column density and finer spacing than in the external filaments.

*2.4. Summary of HFS properties*

Figures 1-15 show that young stellar groups are often associated with "hubs" of high column density, radiating filaments of parsec scale, which have lower density of gas and associated stars. Hubs are often the densest parts of more extended complexes, with peak column density $10^{22-23}$ cm$^{-2}$. In a few cases, hub-filament systems have "neighbor" filaments as in Taurus, with similar elongation and direction. Smaller hubs tend to be relatively round, with relatively fewer stars, lower column density, and a few radiating filaments. Larger hubs tend to be more elongated, with more stars, higher column density, and 5-10 radiating filaments. These filaments tend to be nearly parallel to each other, and are directed along the short axis of the hub, as in Corona Australis and G343.78-0.24.

HFS is seen in dust emission and absorption, and in molecular line emission. It is seen in optical dark clouds within a few hundred pc, and in infrared dark clouds at distances ∼ 3 kpc. In the IRDCs, candidate massive protostars are detected via their luminous mid-infrared emission. Lower-mass clusters are expected to be associated with such massive protostars, but their conclusive detection in IRDCs has not been reported, perhaps due to insufficient sensitivity and resolution.

HFS is much less common in molecular clouds than "filamentary structure," which appears to describe much of the gas in a molecular cloud, throughout its history. HFS is prevalent among young stellar groups, but the definition adopted here for a young stellar group selects only the densest parsec-scale gas in a molecular cloud, and only during the time when it is actively forming stars. Thus the Taurus complex is "filamentary" and contains six groups of young stars, yet only one such group, L1495, has associated HFS.

The geometrical features of HFS can be summarized as the radiation of filaments from denser "hubs," and the tendency for filaments to have similar spacing and direction. The specific nature of these properties suggests that HFS may be a "fingerprint" of a distinct physical process, and is not an artifact of biased selection or small-sample statistics. To better quantify the incidence and properties of HFS, it would be useful to extend the limited survey of nearby dark clouds and IRDCs reported here, to more complete surveys with more uniform sensitivity and resolution.

### 3. Models of filamentary structure

This section describes cloud structure models which have some similarity to observed hub-filament structure, in subsections 3.1-3.2, and presents a new model in subsections 3.3-3.8.

*3.1. Flow-driven filamentary structure*

Filamentary structure in clouds can arise from collisions of randomly directed flows, as in models of hydrodynamic turbulence (Klessen et al 2004, Jappsen et al 2005). However in these models the filaments tend to have irregular spacings and directions. The "nodes" where such filaments intersect are sites of increased star formation, but they do not have sufficient increase in column density and in stellar surface density to match those associated with observed hubs.

Uniform cylinders of gas approaching along a common symmetry axis develop a "splash" pattern of filaments directed radially outward in the collision plane (Vazquez-Semadeni et al 2007, Heitsch et al 2008). This pattern resembles a hub with radial filaments, but its initial state may be unrealistic, since no evidence has been reported of clouds which are either uniform, or colliding.

The formation of protostars in the central part of a cloud can cause outflows to "sculpt" radially directed filaments by removing lower-density gas between them (Li and Nakamura 2006). However, the clumpy distributions of observed protostars, and the irregular directions of their outflows, seem unlikely to produce the similar spacings and directions of the filaments in Figures 1-15.

Hubs and filaments are also seen in the simulations of a magnetized cloud whose turbulence dissipates, causing the cloud to condense along its field lines into a dense layer (Nakamura and Li 2008). Ambipolar diffusion allows gravitational instabilities to structure the layer into hubs and filaments. However these filaments do not have the regular spacing and nearly parallel directions seen in clouds such as Ophiuchus (Figure 8), Corona Australis (Figure 9), G335.43-0.24 (Figure 14), and G345.00-0.22 (Figure 15).

Many authors have suggested that young stellar clusters and their associated dense gas arise from the compression associated with parsec-scale flows, since young clusters are often seen near OB associations and HII regions, whose winds and shocks can sweep up a low-density medium and compress existing condensations (Blaauw 1964, Elmegreen & Lada 1977, de Geus 1992, Wilson et al 2005). This idea motivates the model presented in this paper, since such large-scale compression can condense gas into a layer, and since self-gravitating layers can give rise to the similarly-spaced filaments seen in nearby star-forming regions.

*3.2. Filaments arising from layer fragmentation*

Layer fragmentation has been studied extensively. In a horizontally smooth, self-gravitating isothermal layer (Spitzer 1942), periodic small-amplitude perturbations of sufficiently long wavelength can grow if their form is a circularly symmetric Bessel function or a sinusoid (Ledoux 1951), leading to fragments which resemble circularly symmetric clumps or parallel filaments, respectively. Similar fragmentation can occur in layers of gas which are not isothermal, or which are threaded with nonzero magnetic field (Larson 1985, Nagai et al 1998). A layer bounded by shocks can similarly fragment (Whitworth et al 1994) or can develop a corrugated structure via the nonlinear thin shell instability (Vishniac 1994). These analyses offer insight into how layers respond to perturbations, but do not predict their long-term evolution.

*3.3. Filamentary layer in gravitational equilibrium*

The modulated isothermal layer (Schmid-Burgk 1967) represents the parallel filament systems of Figs 1-15 better than do any of the foregoing models. This subsection outlines the basic model properties.

In the equilibrium model of Schmid-Burgk (1967), the density n of an infinitely extended layer is modulated in one horizontal direction and is uniform in the perpendicular horizontal direction, according to

$$\frac{n}{n_c} = \frac{1-A^2}{\left[\cosh(z/l_M) + A\cos(x/l_M)\right]^2} \quad (1),$$

where the modulation amplitude A ranges from 0 to 1. When A=0 the layer is here called "unmodulated." When A=0 and z=0, equation (1) shows that the central density is equal to $n_c$. In equation (1), $l_M$ is the modulation scale length,

$$l_M \equiv \frac{\sigma}{\sqrt{2\pi G m n_c}} \quad (2)$$

which is equal to the scale height of the unmodulated layer and is dimensionally equal to the Jeans length. Here $\sigma$ is the one-dimensional velocity dispersion, m is the mean molecular mass, and G is the gravitational constant.

When A=0, equation (1) reduces to the isothermal layer of Spitzer (1942). As A→1 the system approaches a periodic series of parallel isothermal cylinders (Stodolkiewicz 1963, Ostriker 1964) whose axes lie in the z=0 plane and have separation $\lambda \equiv 2\pi l_M$. For intermediate values of A, the density has parallel ridges and valleys. States of increasing modulation are energetically favored states of lower total energy (Schmid-Burgk 1967, Curry 2000).

Equation (1) differs trivially from the solution given by Schmid-Burgk (1967) in that the denominator has a plus sign instead of a minus sign. This change shifts the phase of the modulation pattern by half a cycle, which proves to give better agreement with observations, as discussed in Section 4.3.

The modulated equilibrium of an isothermal layer has received little study, despite its possible utility in describing star-forming clouds. It is expected that its dense ridges have a critical mass against fragmentation between those of the smooth isothermal layer and the isothermal cylinder, since these structures have very similar critical mass. The critical mass of the cylinder exceeds that of the layer by less than 20%, for the same temperature and peak density (Larson 1985). The analogous modulated equilibrium of a self-gravitating isothermal cylinder has been described as a model of dense cores in a filamentary cloud (Curry 2000).

*3.4. Column density of a modulated equilibrium layer*

The column density is more useful than the number density for comparison with observed cloud images. The column density of a face-on Schmid-Burgk layer does not appear to have been described previously, so it is presented here. The column density of the modulated layer $N_M$ is obtained by integrating eq. (1) from z=-Z/2 to Z/2, giving

$$N_M = N_U F \tag{3}$$

where $N_U$ is the column density of the unmodulated layer, given by

$$N_U = \sigma \sqrt{\frac{2n_c}{\pi m G}} \tanh \Lambda_l \tag{4}$$

where $\Lambda_l$ is the number of modulation lengths in the half-thickness of the layer,

$$\Lambda_l \equiv \frac{Z}{2l_M} \tag{5}$$

and where F is the column density modulation function,

$$F \equiv \left(\frac{1-A^2}{1-\alpha^2}\right)\left[\frac{1}{1+\alpha\,\text{sech}\Lambda_l} - \frac{2\alpha\coth\Lambda_l}{\sqrt{1-\alpha^2}}\text{Arctan}\left(\frac{1-\alpha}{\sqrt{1-\alpha^2}}\tanh\frac{\Lambda_l}{2}\right)\right] \tag{6}$$

with

$$\alpha \equiv A\cos(x/l_M) \tag{7}.$$

Equations (3)-(7) show that the modulated column density is the product of the unmodulated column density and a periodic function of position in the direction of modulation. In equation (6), F depends only weakly on $\Lambda_l$ and can often be approximated by its limit when $\Lambda_l \to \infty$,

$$F_\infty \equiv \left(\frac{1-A^2}{1-\alpha^2}\right)\left(1 - \frac{2\alpha}{\sqrt{1-\alpha^2}} \text{Arctan} \frac{1-\alpha}{\sqrt{1-\alpha^2}}\right) \qquad (8).$$

Equation (8) shows that $F_\infty$ has maximum values $F_+$ when $x/(2\pi l_M)$ is an even integer and minimum values $F_-$ when $x/(2\pi l_M)$ is an odd integer, and that these extrema are given by

$$F_\pm = 1 \pm \frac{2A}{\sqrt{1-A^2}} \text{Arctan} \frac{1 \pm A}{\sqrt{1-A^2}} \qquad (9).$$

For a typical value A=0.34, to be considered in Section 4, $F_- = 0.55$ and $F_+ = 1.70$, so that the ridge-valley column density ratio is $F_+/F_- = 3.0$.

The foregoing modulated layer can describe systems of parallel filaments, because it has finite amplitude, unlike the infinitesimally small amplitude of perturbation analysis (Schmid-Burgk 1967, Curry 2000). Also, the modulated layer is an energetically favored equilibrium, so configurations which come close to its structure can persist. The flows, shocks and instabilities discussed above may help to generate such configurations.

As an initial state, the constant column density in equation (4) is too uniform to match the low-density gas observed in and near star-forming regions. Also, the coresponding modulated structure gives exactly parallel filaments but no hubs. The following subsections describe a more realistic initial state and modifications of the Schmid-Burgk model, which together give a better match to observed hub-filament structure.

*3.5. Compressed Clumpy Medium Model*

As noted above, the original Schmid-Burgk layer does not match observed clouds, because its filaments are perfectly parallel and do not converge onto hubs. The paper therefore seeks a modification of the Schmid-Burgk layer which has converging filaments. Due to this modification, the layer is not in equilibrium, but approaches equilibrium in its outer parts. It can still be a useful model provided its appearance survives long enough to be observable, as discussed in Section 5.6.

The origin of this modified system is the compression of a nonuniform initial medium, as described in Section 3.6. This nonuniform medium is a "clump" embedded in uniform gas. This combination is more realistic than a perfectly uniform medium, and after compression the clump becomes the hub to which the Schmid-Burgk filaments converge.

This initial state is near equilibrium but not in equilibrium. To be realistic the initial clumpy medium should last long enough to be compressed, but it need not be in exact force balance. The initial clumpy medium is modeled as the sum of the densities of an isothermal equilibrium spheroid and of a uniform background. The density functions are constrained to match observations. Their small departure from equilibrium is discussed in an earlier paper (Myers 2008).

This initial clumpy medium is compressed until it becomes self-gravitating in its direction of compression. Expressions for the layer column density are given in the Appendix. After 1D compression the unmodulated column density is simply the column density of a segment of the initial clumpy medium.

The modulated compressed layer approaches exact 3D equilibrium in the limits where the initial clump is vanishingly small or infinitely distant. The innermost part of the layer departs most from equilibrium and contracts horizontally toward the center of the initial clump. This collapsing central region is expected to have the greatest efficiency of star formation, as indicated by the observations in Section 2, and as discussed in Section 5.6.

*3.6. Initial medium model*

Sensitive, high-resolution observations of the low-density interstellar gas generally show clumpy and filamentary structure, even in gas which appeared relatively featureless in earlier observations of lower quality (Goldsmith et al 2008). Thus in this paper, the initial state is an idealized version of a clumpy filamentary medium. It consists of a prolate isothermal spheroid embedded in an uniform medium of lower density, similar to the "IS+U" model of Myers (2008).

The spheroid is chosen to resemble observed low-density "clumps" in nearby molecular clouds (Kirk, Johnstone & Di Francesco 2006, Bergin & Tafalla 2007) and the medium is chosen to resemble their surrounding lower-density gas. When the spheroid has low aspect ratio, the medium is "clumpy." When the spheroid is elongated, the medium is "filamentary." A more realistic model would have more than one such spheroid, but an initial state of one spheroid in an extended medium is sufficient to reproduce the main features of hub-filament structure.

The prolate spheroid model is a "stretched" version of an isothermal equilibrium sphere, and is a simple way to describe observed elongated clumps. Its physical basis is limited because it requires anisotropic pressure to maintain its equilibrium configuration. A more realistic model of elongated isothermal equilibrium with central condensation is available (Curry 2000), but is more difficult to implement. The main features of the modulated structure do not depend on differences between these initial models, so for simplicity the prolate spheroid is adopted here.

Expressions for the density and column density of the initial clumpy medium are given in the Appendix.

*3.7. Modulated Column Density of Compressed Medium*

As discussed above, compression of a clumpy medium provides a layer whose outer region can host Schmid-Burgk filaments, and whose inner region has the high column density and efficient star formation seen in the hubs described in Section 2. This subsection motivates the calculation of the modulated column density of this layer.

The Schmid-Burgk layer has modulated column density $N_M = N_U F_\infty$ as in equation (3), where $N_U$ is the column density of the unmodulated layer. For the pure Schmid-Burgk layer, $N_U$ is equal to $N_u$, the column density of the vertically self-gravitating, horizontally uniform, isothermal layer of Spitzer (1942). In contrast the compressed clumpy medium is modelled as vertically self-gravitating, but it is not horizontally uniform: its unmodulated column density $N_U$ is the sum of the uniform and spheroidal components, $N_U = N_u + N_s$ as in equation (A8).

As a consequence, the modulated column density of a horizontally nonuniform layer can be approximated from the Schmid-Burgk column density in equation (3) only if the density profile of the layer is sufficiently similar to the Schmid-Burgk density in equation (2). The simplest way to meet this condition is to require that at all positions, the density $n_s$ of the spheroidal component is significantly less than the density $n_u$ of the horizontally uniform component, i.e. $n_s \ll n_u$.

There is no obvious way to satisfy this condition $n_s \ll n_u$ at all positions while also matching the observational requirement that the peak hub column density significantly exceed that of its associated filaments, i.e. $N_{s,max} \gg N_u$. These two conditions could be met in principle if the velocity dispersion associated with the spheroidal component greatly exceeded that of the horizontally uniform component. But it is difficult to understand how such a difference in velocity dispersions could arise and persist.

Thus the Schmid-Burgk equilibrium of a pure filamentary layer cannot approximate a hub-filament layer, except in its outermost parts.

As an alternative, a layer is described here which approaches the Schmid-Burgk equilibrium in the limit of large radial distance from the center, while its centrally condensed interior is radially collapsing. This situation cannot be modelled analytically as exactly as a pure equilibrium, but it may correspond to the observed central regions of high star formation efficiency, surrounded by filamentary extensions.

In this picture, the unmodulated column density arises from compression of the initial clumpy medium. Its modulation can be calculated reliably only at large radius where the unmodulated layer approaches horizontal uniformity. Such a region may begin where the total unmodulated column density $N_U$ exceeds the horizontally uniform component $N_u$ by a small factor, for example by the factor 1.1. For the example to be calculated in Section 4, the corresponding radius is 4.7 pc. Beyond this radius the modulated column density should approach $N_M = N_u F_\infty$ as in equation (3) for the pure Schmid-Burgk layer.

At smaller radii where the nonuniform spheroidal component becomes more important, no guide to the modulated structure come from the physics of Schmid-Burgk equilbrium, since the necessary equilibrium conditions do not apply. On the other hand, some observed hubs show parallel internal filaments, as discussed in Section 2.3, and a numerical study of a layer created by collision of uniform clouds shows filamentary structure which persists during much of its collapse (Vazquez-Semadeni et al 2007).

Therefore, in an attempt to describe hub-filament systems, two cases have been calculated whose modulation properties are expected to bracket observed systems. Each reflects the general idea that a vertically self-gravitating layer can support filaments, and that their spacing is proportional to the scale height of the layer. At large radius, they approximate the Schmid-Burgk equilibrium. At smaller radius, they represent extrapolations of the Schmid-Burgk equilibrium beyond its physically justified regime.

In the case of "minimum modulation," the horizontally uniform component of the layer has the Schmid-Burgk modulation of an isolated layer, while the spheroidal component has no modulation. The minimum modulated column density is given by

$$N_{M\,min} = N_s + N_u F \qquad (10)$$

where the modulation function F is calculated using the modulation scale length of equation (2) based only on the constant central density of the horizontally uniform component.

In the case of "maximum modulation" the Schmid-Burgk modulation is applied to the total unmodulated column density:

$$N_{M\,max} = (N_s + N_u)F \qquad (11).$$

Here F is calculated with the same amplitude A as in the minimum case, and the modulation scale length is based on the varying central density due to both components of the layer.

These cases of minimum and maximum modulation differ sharply in their resemblance to observed hub-filament systems. Both show parallel filamentary lanes at large distance from the hub. But as neighboring filaments approach the hub, with minimum modulation they diverge and follow the outer contours of the hub. With maximum modulation, the same contours converge and join the hub, more closely matching the hub-filament structure described in Section 2.

These different spatial relations of filaments to the hub arise because the modulation scale length decreases with increasing central density, as in equation (2). In the minimum modulation case the central density is constant, while in the maximum modulation case the central density increases inward, shrinking the scale of the filamentary pattern. It is also possible that the convergence of filaments onto the hub reflects gravitational infall along the filaments, as discussed in Section 5.5.

Therefore in Section 4 the column density structure of the compressed clumpy medium will be presented using eq. (11) representing the "maximum modulation" case. Although it is based on an extrapolation of the physically justified Schmid-Burgk solution, this procedure provides a much closer match to observed HFS than does the alternate case of minimum modulation.

If the layer extends from the midplane for more than about a scale height, the column density modulation function F in equation (6) may be approximated by the simpler expression for an infinite layer, $F_\infty$ in equation (8). For the parameters to be adopted in Section 4, the approximation $F \approx F_\infty$ is better than 5% for all positions in the layer. This is an acceptable level of accuracy for this calculation, so henceforth $F_\infty$ will be used for F.

## 4. Modulated layer column density

This section uses observed cloud properties based on Figures 1-9 and Table 1 to set model parameter values, and presents contour maps of column density obtained from the adopted parameters and the model equations of Section 3.

*4.1. Input parameters*

The maximum spacing between adjacent filaments, $\lambda_{max}$, is taken to be 5.0 pc, based on the well-defined spacings in Taurus (Figure 1) and Ophiuchus (Figure 8). Clouds with smaller filament spacings can be matched by use of the scaling relations in section 4.4. The peak initial density $n_0$ is taken as 5000 $cm^{-3}$, close to values estimated for "clumps" in star-forming clouds based on observations of emission in the J=1-0 line of $^{13}CO$ and on star-count extinction (Kirk, Johnstone & Di Francesco 2006, Bergin & Tafalla 2007). The ratio of modulated peak to valley column density $F_+/F_-$ is taken to be 3, based on the extinction study of three filaments in Taurus (Arce & Goodman 1999).

Observations constrain estimates of the initial uniform density $n_u$, the initial uniform component of column density $N_u$, and the accumulation length $L = N_u/n_u$. The mean density of nearby molecular clouds is 300-600 $cm^{-3}$ when the cloud boundary is set by extinction $A_V = 2$ magnitudes (Evans et al 2008). A representative value for $n_u$ may lie below this range, since CO observations indicate that more than half the molecular mass of the Taurus complex lies below $A_V=2$ (Goldsmith et al 2008). The initial gas before compression should extend similarly in the direction of compression and in the plane of the sky, so it is expected that L = 5-10 pc. The initial uniform column density should be comparable to the "background" component of column density in nearby clouds, typically 1-3 × $10^{21}$ cm$^{-2}$. These constraints are satisfied by the parameter values $n_u$=200 cm$^{-3}$, $N_u$ = 3 × $10^{21}$ cm$^{-2}$, and L=5 pc, which are adopted here.

The velocity dispersion σ is obtained from the maximum filament spacing $\lambda_{max}$ by evaluating the modulation scale length $l_{Mmax}$ for β >> 1, where the spheroid contribution to the unmodulated layer density is negligible. Then in eqs. (A14) and (A16), $n_{min} \approx n_u$ and $N_U \approx n_u L$, yielding a quadratic equation for the central density $n_c$, with solution

$$n_c = \frac{n_u}{2}\left[1 + \sqrt{1 + (L/l_{M\,max})^2}\right] \quad (12).$$

Combining equations (A14) and (A16) with $\lambda_{max}=2\pi\, l_{Mmax}$ as in Section 3.4 gives

$$\sigma^2 = \frac{mn_u G \lambda_{max}^2}{4\pi}\left[1 + \sqrt{1 + (2\pi L/\lambda_{max})^2}\right] \quad (13).$$

For the foregoing values of $n_u$, L, and $\lambda_{max}$, equation (13) gives σ=0.84 km s$^{-1}$. This value corresponds to observed velocity dispersions in lines which trace parsec-scale structure in nearby clouds, such as the J=1-0 line of $^{13}$CO (e.g. Ridge et al 2006). However it exceeds the thermal velocity dispersion in these same regions by a factor ~4. The supersonic nature of the velocity dispersion is discussed in Section 5.

The modulation amplitude A = 0.34 is obtained from equation (9) for the adopted column density ratio $F_+/F_- = 3.0$.

The final parameter is the initial elongation g in equation (A5), which is set to 1 and to 3 to give initial clumps which are spherical and elongated, leading to hubs whose projected shapes are respectively round and elongated as seen in Figures 1-15. Thus all of the six independent parameters are either set or constrained by observations, and these choices completely specify the model. They are summarized in Table 2.

With the input parameters of Table 2 and the model equations of Section 3, the horizontal structure of the unmodulated column density $N_U$ and the modulated column density $N_M$ were calculated for layers compressed from clumpy media having spherical and elongated initial clumps. The calculations extend from x=-20 pc to x=20 pc in steps of 0.01 pc and from y=-10 pc to 10 pc, in steps of 0.4 pc for |y| > 4 pc and in steps of 0.2 pc for |y| < 4 pc. The results are presented in the next two subsections.

*4.2. Spherical initial clump*

Figure 16 shows contours of unmodulated column density $N_U$ for the adopted parameters in Table 2 when the initial clump is spherically symmetric, i.e. when its elongation parameter is g=1. The lowest contour, at $3.2 \times 10^{21}$ cm$^{-2}$, is set slightly above the uniform column density level, $N_u = 3.0 \times 10^{21}$ cm$^{-2}$. This choice shows the transition between the nearly uniform outer zone and the centrally condensed inner zone, by the large spacing of the outermost contours and the progressively smaller spacing of the inner contours. The masses within the central 0.5 pc and 5.0 pc are respectively 250 $M_O$ and 6100 $M_O$, in reasonable accord with estimates in nearby clouds (Jørgensen et al 2008, Evans et al 2008).

Figure 17 shows contours of modulated column density for the same system as in Figure 16, with the adopted modulation parameter A=0.34, corresponding to peak-to-valley column density ratio of 3.0. The column density pattern shows both the one-dimensional modulation of the Schmid-Burgk equilibrium and the circularly symmetric radial structure of the initial system. The structure resembles the HFS summarized in Section 2. It is highly "filamentary," with the few most prominent filaments converging on a central "hub" of high column density, as is seen most clearly in Figures 2-4, 6, and 11-13. The hub has two column density peaks separated by 1 pc. These peaks are continuations of the large-scale filaments spaced by 5 pc, seen where they have converged to their point of closest approach. Their small central spacing reflects the fact that the modulation length $l_M$ of the unmodulated midplane gas is much smaller at the center than at the outside.

The hub and radiating filaments in Figure 17 are accompanied by parallel neighbor filaments of lower column density, as seen in Figures 2, 8, and 13. The minimum contour is set to 5.25 × $10^{21}$ cm$^{-2}$, low enough to reveal the neighbor filaments but high enough to show that they are less extended than the more prominent central filaments.

The structure of Figure 17 is based on a choice of modulation phase. When the unmodulated gas is uniform, the phase of this dependence is arbitrary, as noted in Section 2. However when the unmodulated gas is centrally condensed around (x, y) = 0, the choice of phase affects the spatial relation of the filaments to the hub. In the choice adopted here, filaments have even symmetry about the x=0 line, and the most prominent filaments appear as "doublets" at $x=\pm l_M/\pi$. This appearance agrees better with most of the clouds in Figures 1-15 than when the sign in the denominator of equation (1) is minus instead of plus. In that case the filaments have odd symmetry about x = 0, and the most prominent filament is a "singlet" at x = 0.

*4.3. Elongated initial clump*

Figure 18 gives the column density structure of the clumpy medium whose initial spheroid is elongated by a factor g=3. This initial structure is motivated by the observation of elongated clumps in molecular cloud maps, and is a horizontally stretched version of that in Figure 16. As before, the wide spacing of the outermost contours indicates the transition between the uniform and spheroidal zones of the initial clumpy medium.

Figure 19 shows the column density of the modulated layer whose initial medium has an elongated clump. The pattern shows Schmid-Burgk modulation in the x direction in the form of nearly parallel filaments. These converge toward an elongated dense hub, whose shape reflects the elongation of the initial clump. This same elongation is seen in the outer boundary of the filaments. These features are similar to those in Figure 17 for the modulated layer whose initial medium has a spherical clump. However because of the increased initial elongation, many more filaments are visible than in Figure 17, and the similarity of the filament directions and spacings is much more evident. The structure in Figure 19 resembles observed clouds with many filaments, shown in Figures 8, 9, 14, and 15.

The structure of Figure 19 is based on a choice of modulation direction. When the initial clump is spherical, the direction of modulation, represented here as the x-direction, is arbitrary. However when the initial clump is elongated, the direction of modulation with respect to the direction of elongation affects the modulated structure. In the following calculation the modulation direction is chosen to align with the principal axis of the initial clump, to give better agreement with the clouds in Figures 1-15. The physical basis of this alignment is probably set during the initial compression, and would be a useful subject for numerical studies.

*4.4. Pattern scaling*

The spatial patterns of column density in Figures 17 and 19 can scale by changing their parameters together in particular ways. A measure of the pattern shape is the ratio $\lambda_{max}/a$ of the maximum filament spacing and the spheroid scale length, which is proportional to the number of filaments per hub. From equations (2) and (A7),

$$\left(\frac{\lambda_{max}}{a}\right)^2 = 16\pi^2 \left(\frac{n_0}{n_u} - 1\right)\left[1 + \sqrt{1 + \left(\frac{2\pi L}{\lambda_{max}}\right)^2}\right]^{-1} \quad (14).$$

Thus if the same pattern is present in two clouds, denoted with subscripts 1 and 2, they must follow

$$1 = \frac{[(n_0/n_u) - 1]_2}{[(n_0/n_u) - 1]_1} \frac{\left[1 + \sqrt{1 + (2\pi L/\lambda_{max})^2}\right]_2}{\left[1 + \sqrt{1 + (2\pi L/\lambda_{max})^2}\right]_1} \quad (15).$$

A simple way to satisfy eq. (15) is to require that each of the two multiplying terms on the right side be unity, so that

$$\left(\frac{n_0}{n_u}\right)_1 = \left(\frac{n_0}{n_u}\right)_2 \tag{16}$$

and

$$\left(\frac{L}{\lambda_{max}}\right)_1 = \left(\frac{L}{\lambda_{max}}\right)_2 \tag{17}.$$

Then equations (14), (15) and (17) give scaling relations for two clouds which have the same hub-filament pattern shape and a size ratio $s \equiv \lambda_{max1}/\lambda_{max2}$:

$$\frac{N_{max1}}{N_{max2}} = \frac{N_{u1}}{N_{u2}} = \frac{n_{01}}{n_{02}}s = \frac{n_{u1}}{n_{u2}}s \tag{18}$$

and

$$\frac{\sigma_1}{\sigma_2} = \left(\frac{n_{01}}{n_{02}}\right)^{1/2} s = \left(\frac{n_{u1}}{n_{u2}}\right)^{1/2} s \tag{19}.$$

Equations (18) and (19) show how initial parameters must relate to reproduce a given pattern shape at the same size scale or at a different size scale. For example if an infrared dark cloud had the same pattern and size scale as in Figure 19 but peak column density greater by a factor 5, it would require a denser and hotter initial medium, with initial values $N_u$, $n_0$, and $n_u$ greater by a factor 5 and $\sigma$ greater by a factor $\sqrt{5}$. Alternately if a nearby dark cloud had the same pattern and peak column density as in Figure 17 but size scale smaller by a factor 5, it would require a denser and cooler initial medium, with column density $N_u$ unchanged, initial densities $n_0$ and $n_u$ greater by a factor 5, and velocity dispersion $\sigma$ less by a factor $\sqrt{5}$. Evidently a variety of initial conditions can lead to the filamentary patterns in Figures 17 and 19.

## 5. Discussion

This section discusses the observational evidence for hub-filament structure, the physical basis of the models in Sections 3 and 4, and some implications for our understanding of groups and clusters. A theme common to several topics in the section is the need for numerical simulations of hub-filament systems. Many aspects of hub-filament structure involve dynamical processes which cannot be described by the simple analytic models in this paper. While the analytic models can offer physical insight for limiting cases, numerical simulations are needed to make further progress on development of filaments following compression, their spatial relation to the central hub, and on survival of the hub-filament structure over time.

### 5.1. Observed hub-filament structure

The images in Figures 1-15 are based on a literature survey and so do not have uniform sensitivity, resolution, or areal coverage. They also are difficult to compare with model images because they are projected images of 3D objects which are unlikely to have planar symmetry. Those whose geometry is flattened may not lie in the plane of the sky. Nonetheless, a significant number of examples show HFS, and for plausible ways to define the sample, a significant fraction of the nearest young stellar groups have associated HFS. Although star-forming clouds have long been known to be "filamentary" the new feature reported here is that of multiple filaments radiating out from a central hub. This property has not been described before, and it is more evident now because recent observations at optical, infrared, and millimeter wavelengths are more sensitive and cover a much greater area than in the recent past. Such multiple filaments are important because they motivate models of modulated layers, while single filaments do not.

*5.2. Compression*

Parsec-scale regions of cloud gas can be compressed by large-scale flows (Vazquez-Semadeni et al 2007), or by stellar wind bubbles, expanding H II regions, or supernova remnants (Whitworth 2007, Elmegreen 1998). A detailed model of the origin of the Ophiuchus complex suggests that massive stars in the Upper Scorpius and Upper-Centaurus Lupus subgroups originated wind-blown bubbles whose expansion swept up ambient gas to form the complex seen today (de Geus 1992). In this model OB winds are combined with supernova explosions help to inflate the bubbles. Similarly it has been suggested that both the Ophiuchus and Lupus complexes may have formed as a consequence of the winds, ionization and supernova shells from these subgroups of the Sco OB 2 association (Tachihara et al 2001, Figures 10 and 11). The molecular clouds of the Cyg X complex appear to have been shaped by winds and radiation from the OB stars in the Cyg OB 2 association (Schneider et al 2006). Similarly it was suggested that the kinematics and structure of the CO gas in Orion A and B are consistent with stellar wind driven compression centered on Ori OB 1b (Wilson et al 2005). Thus many nearby star-forming regions appear to be close enough to OB subgroups to account for their compression by OB winds. Other sources of compression such as colliding turbulent flows and expanding H II regions may also play a role, but proximity to OB associations seems to have the greatest observational support.

*5.3. Equilibration time*

The time for a suddenly compressed layer to become vertically self-gravitating is of order the free-fall time of the initial uncompressed medium (Whitworth 1994). Thus interstellar gas with uniform density of a few $cm^{-3}$ takes a few tens of Myr to become self-gravitating after sudden compression. This equilibration time is much longer than the typical star-forming duration of a few Myr. If star-forming fragmentation required the hosting layer to first be self-gravitating, most dense layers would be starless, in contrast to observation (Larson 2005). Thus to match observations, the gas to be compressed must be sufficiently dense, or the compressed layer must form stars before it is fully self-gravitating.

The clumpy initial medium considered here is centrally condensed, so its equilibration time is shorter at small radius than at large radius. For the medium shown in Figure 16, the free-fall time is 0.6 Myr at radius 0.5 pc and 2.1 Myr at 5 pc. Thus the densest hub gas can fragment and begin to form stars while its surrounding gas is still condensing toward an equilibrium configuration. If radiating filaments arise from growth of gravitational instabilities in an equilibrium layer, they might not appear until 1-2 Myr after the first central stars. If instead both hubs and filaments arise promptly during the compression which forms the layer (Vazquez-Semadeni et al 2007, Heitsch et al 2008), the role of self-gravity might be to selectively preserve the ridges and valleys which most closely resemble those of modulated equilibrium.

*5.4. Supersonic motions*

The initial medium and the compressed layer models of Sections 3 and 4 are based on self-gravity and isotropic thermal pressure. To match observed column densities over parsec scale lengths, these models require that the velocity dispersion exceed by a factor of several the thermal velocity dispersion known from estimates of the gas temperature. Such supersonic motions are also needed to match line widths observed on these scales (e.g. Ridge et al 2006). Thus the models discussed here resemble many prior equilibrium models of gravity and isotropic pressure, in their use of supersonic velocity dispersion to match observations on parsec scales (Larson 1981, Myers & Fuller 1992, McLaughlin & Pudritz 1996, McKee & Tan 2003).

A straightforward way to reconcile these results is to posit that "microturbulent" motions provide a nonthermal component of isotropic pressure. Hoewever, observations of magnetic fields and simulations of magnetic turbulence indicate that that its motions are anisotropic and highly dissipative. This description of turbulence is not analogous to thermal pressure in its support against self-gravity (MacLow & Klessen 2004). Thus, understanding the nature of supersonic motions in interstellar clouds remains a challenge. Part of this challenge is to understand why isothermal models with supersonic velocity dispersion such as that in Sections 3 and 4 successfully describe large-scale features of molecular clouds.

The models in this paper are isothermal, so the velocity dispersion $\sigma$ has a single value over all positions and times. This simplification contradicts substantial evidence from molecular line observations for increasing velocity dispersion with scale size, especially on parsec scales (Larson 1981, Goodman et al 1998). Furthermore it is expected that after gas is condensed by compression, it will cool to lower temperature and velocity dispersion rather than remaining the same, as assumed here. The justification for such a simple treatment of velocity dispersion is that the modulated equilibrium solution has so far been described only for an isothermal system (Schmid-Burgk 1967). It would be useful to further investigate modulated layer equilibria for other equations of state, and for unmodulated gas which is not horizontally uniform.

*5.5. Convergence of filaments onto hubs*

The convergence of filamentary arms onto hubs in the model of Section 3 is based on the idea that filament spacing scales with the scale height of the layer, which decreases inward. As noted in Section 3.7 this idea represents an extension of the Schmid-Burgk model where the scale height is constant. This extension may be valid as long as the scale height varies sufficiently slowly with position. However it is also possible that the observed convergence of filamentary arms onto hubs has a more dynamic basis, such as gravitational inflow of filament gas toward the attracting hub (e.g. Balsara, Ward-Thompson & Crutcher 2001, Banerjee & Pudritz 2007). Simulations of the compression of a clumpy medium may help to resolve this issue.

*5.6. Survival of HFS*

The layer formed by compression of a clumpy medium, described in Sections 3 and 4, is centrally condensed with surface density declining outward as $\beta^{-1}$, resembling a Mestel disk (Mestel 1963) bounded by zones of uniform column density at small and large radius. On large scales, this layer is out of radial equilibrium, and so it should contract, fragment and form stars within a free fall time of ~2 Myr, unless it is slowed by rotation or by a sufficient gradient in magnetic pressure.

The initial structure of a layer may be substantially altered during such contraction. A Mestel disk profile becomes steeper on the inside than on the outside (Proszkow & Myers 2008). An initial surface density profile with a sharp edge generates more internal structure during contraction than does a profile with a tapered edge (Burkert & Hartmann 2004). An initially ellipsoidal layer contracts fastest along its short axis, increasing its eccentricity as in the collapse of aspherical 3D bodies (Lin, Mestel & Shu 1965).

Nonetheless, once hub-filament structure is established in a layer, its general appearance may remain detectable during much of the layer contraction. The innermost parts of the layer have the densest gas, the shortest free-fall time, and thus the greatest likelihood of fragmentation and star formation. The high incidence of young stellar groups in hubs, discussed in Section 2, supports this picture. Therefore the detailed structure of the hub region of the model should be rapidly overtaken by these dynamical effects. On the other hand, the outermost parts of the layer should contract much more slowly and can be expected to preserve their initial filamentary structure.

A lower limit on the "smoothing time" of neighboring filaments is the time for gas to travel the ridge-valley distance at the sound speed. This time is 3.0 Myr for the large-scale spacing 5.0 pc in Figures 17 and 20, whereas the free fall time of the same system is everywhere less than 2.4 Myr. Thus filamentary structure can not smooth out during large-scale gravitational collapse. This estimate corroborates the filament evolution seen in the simulations of colliding cylinders by Vazquez-Semadeni et al (2007). There, gas in the collision plane contracts radially inward, and does not destroy the filamentary structure until nearly all of the gas has reached the center.

In some cases the survival of HFS may be limited sooner by smaller-scale effects. These include hub dispersal due to winds, outflows, and heating from embedded young stars, and fragmentation of filaments into isolated cores. In the examples of Section 2, one may identify NGC 1333 as a region where the hub has begun its dispersal and IC348 where the dispersal is nearly complete, over a time of a few Myr (Muench et al 2007). Nonetheless, the high incidence of HFS described in Section 2 suggests that such small-scale agents of dispersal do not generally destroy HFS during the time when it is associated with young clusters.

*5.7. Implications*

The observations and models presented here suggest that HFS may be a "fingerprint" of the compression which forms and structures molecular clouds. If so, it may help explain why molecular clouds are so filamentary: they are formed by compression of low-density gas into modulated layers. It may further explain why cluster-forming regions are more efficient at forming stars than is their surrounding gas: cluster-forming regions are compressed clumps and so have greater density than their surrounding gas. In this picture HFS cannot originate by compression of a purely uniform medium, because that would produce only nearly parallel filaments, and no hubs. Similarly HFS cannot originate by compression of a centrally condensed clump, because that would produce a hub but no nearly parallel filaments. Only with the combination of both clump and medium does the characteristic HFS pattern arise.

The parallel directions and regular spacing of HFS filaments are hard to reconcile with most models of turbulent origin of molecular clouds because the turbulent flows which make filaments have random directions. Only if one such flow is dominant in sweeping up ambient clumpy gas can one expect the regularity of filaments seen in clouds such as Corona Australis (Figure 9), the Orion integral filament (Figure 10), and the infrared dark clouds G335.43-0.24 (Figure 14) and G345.00-0.22 (Figure 15).

**6. Summary**

The main points of this paper are:

1. Recent observations indicate that young stellar groups are associated with "hubs" or parsec-scale regions of high column density. These hubs radiate filaments having lower column density and fewer stars. Hubs of low aspect ratio have few radiating filaments, while elongated hubs have more filaments. Filaments associated with elongated hubs tend to have regular spacing and extend parallel to the hub minor axis. Some hub-filaments systems seem to be part of a larger pattern of nearly parallel filaments. Figures 1-15 shows that these properties are associated with all nine of the nearest young stellar groups, and are also seen in some clouds forming more massive stars.

2. Most available models of filamentary clouds produce filaments but not the observed hub-filament structure. Those models with hubs radiating filaments do not show the observed regularity of filament spacing and parallel directions.

3. A new model is based on the idea that a self-gravitating layer can develop and maintain filamentary stucture. The initial state is an idealized version of the low-density gas in a molecular cloud. A clump in a uniform medium is compressed into a self-gravitating, centrally condensed layer. The compression may be driven by winds from nearby OB associations. Far from its center, the outer layer undergoes slow radial contraction and remains close to equilibrium, while the interior collapses more rapidly and forms a stellar group with high efficiency.

4. Filamentary structure may be initiated quickly by shocks or splashes associated with the compression, or more slowly by gravitational instabilities. The modulated equilibrium of Schmid-Burgk (1967) is used to approximate the longer-lived structure of the outer layer, and is extrapolated in an attempt to describe its inner modulation.

5. When the modulation scale length is based on the uniform component of the midplane density, parallel filaments diverge as they approach the hub. When instead the scale length is based on the total midplane density, parallel filaments converge onto the hub, giving a better match to observed hub-filament structure. This latter method is used for further calculations.

6. The modulated structures presented here should be considered only as suggestive in the hub regions, and more nearly predictive at large radius as their filaments become parallel. Nonetheless these results may be useful to spur more detailed studies of modulated layers in star-forming regions.

7. The initial parameters of clump density, medium density, and velocity dispersion are chosen to match those of clumps and clouds in nearby star-forming regions. The initial clump is an isothermal spheroid whose peak density is much greater than that of its surrounding uniform medium. This idealized clumpy medium is slightly out of equilibrium but is expected to last long enough to allow compression into a self-gravitating layer.

8. When the initial clump is spherical the calculated modulation pattern shows few radiating filaments, matching the most frequently observed case. When the initial clump is elongated the pattern shows more filaments, and their parallel direction and regular spacing is more apparent, as seen in Corona Australis, Orion, G335.43-0.24, and G345.00-0.22.

9. For a given hub-filament pattern shape, the pattern size follows simple scaling laws with the initial densities, column densities, and velocity dispersion.

10. The filamentary structure observed in molecular clouds may be a "fingerprint" of the compression which structures molecular clouds into modulated layers.

**Acknowledgements**

This paper has benefited from the encouragement and support of Irwin Shapiro, and from helpful discussions with Fred Adams, Lori Allen, Andi Burkert, Paola Caselli, Neal Evans, Gary Fuller, Alyssa Goodman, Rob Gutermuth, Lee Hartmann, Tom Hartquist, Tom Megeath, and Quizhou Zhang. Gus Muench kindly modified a published image of IC348 for use here as Figure 5. An anonymous referee suggested several helpful clarifications of the physical explanation of the procedures used.

**Appendix. Density and column density of initial medium**

The initial density $n_{su}$ is given by

$$n_{su} = n_s + n_u \qquad (A1)$$

where $n_s$ and $n_u$ are respectively the densities of the spheroidal and uniform components. The spheroidal component is an elongated version of the isothermal sphere, and extends the approximation of Natarajan and Lynden-Bell (1997),

$$n_s = (n_0 - n_u)\left(\frac{C}{\gamma^2 + \zeta^2} - \frac{D}{\delta^2 + \zeta^2}\right) \quad (A2)$$

where $n_0$ is the peak density. The constants are C=50 and D=48, and the dimensionless coordinates $\gamma$, $\delta$, and $\zeta$ are related to the cartesian coordinates x, y, and z by

$$\gamma^2 \equiv c^2 + \beta^2 \quad (A3)$$

$$\delta^2 \equiv d^2 + \beta^2 \quad (A4)$$

where $c^2$=10 and $d^2$=12, and where

$$\beta^2 \equiv \left[(x/g)^2 + y^2\right]/a^2 \quad (A5)$$

with

$$\zeta \equiv z/a \quad (A6).$$

In equation (A5) g is the aspect ratio of the prolate spheroid, with g ≥1, and in equations (A5) and (A6) a is the spheroid scale length, given by

$$a^2 \equiv \frac{\sigma^2}{4\pi mG(n_0 - n_u)} \qquad (A7)$$

with constant velocity dispersion σ. Here the line of sight is parallel to the z axis and the long axis of the spheroid lies in the x direction in the plane of the sky.

The uniform component $n_u$ extends indefinitely in each direction. Note that the lower-case subscript "u" means "uniform" while the upper-case subscript "U" in equations (3) and (4) means "unmodulated."

It is assumed that a finite accumulation length L of the initial clumpy medium is compressed along the z direction into a self-gravitating, pressure-truncated layer of constant vertical extent L′ < L. At each horizontal position, the compressed gas has the same column density within L′ as does the initial gas within L. Then the unmodulated column density can be written as the sum of uniform and spheroidal components

$$N_U = N_u + N_s \qquad (A8)$$

where the uniform component is

$$N_u = n_u L \qquad (A9)$$

and the spheroidal component is

$$N_s = \sigma \sqrt{\frac{(n_0 - n_u)}{\pi mG}} \left( \frac{C}{\gamma} \text{Arctan} \frac{L}{2a\gamma} - \frac{D}{\delta} \text{Arctan} \frac{L}{2a\delta} \right) \qquad (A10)$$

obtained by integrating eq. (A2) from z=-L/2 to L/2.

The column density is the sum of these component values,

$$N_U = N_u \left[ 1 + \frac{n_0 - n_u}{n_u \Lambda_s} \left( \frac{C}{\gamma} \text{Arctan} \frac{\Lambda_s}{\gamma} - \frac{D}{\delta} \text{Arctan} \frac{\Lambda_s}{\delta} \right) \right] \quad (A11)$$

where $\Lambda_s$ is the number of spheroid scale lengths contained within half the accumulation length,

$$\Lambda_s \equiv \frac{L}{2a} \quad (A12),$$

defined similarly to equation (5). Equation (A11) shows that if $n_u \ll n_0$, $N_U/N_u$ has a high central "plateau" at the approximate value $\pi n_0/n_u$ when $\beta \ll c$, and a low surrounding plateau at the value 1 when $\beta \gg (n_0/n_u)^{1/2}$. At intermediate values of $\beta$, $N_U$ decreases approximately as $1/\beta$.

At each horizontal position the layer central density $n_c$ is related to the column density $N_U$ and to the density at the layer boundary height $z=L'/2$ by

$$n_c = n(L'/2) + n_{c\infty} \quad (A13)$$

where $n_{c\infty}$ is the central density of an infinitely extended isothermal layer having the same velocity dispersion and column density,

$$n_{c\infty} \equiv \frac{\pi m G N_U^2}{2\sigma^2} \qquad (A14)$$

(Elmegreen & Elmegreen 1978). It is further assumed that gas at the boundary of the compressed zone has the least possible condensation. Then at each horizontal position the density at z=L′/2 is the same as the initial density at z=L/2, obtained from equation (A1). These densities are denoted $n_{min}$, i.e.

$$n_{su}(L/2) = n(L'/2) \equiv n_{min} \qquad (A15).$$

Thus the midplane density of the unmodulated layer is expressed in terms of the initial clumpy medium by

$$n_c = n_{min} + n_{c\infty} \qquad (A16).$$

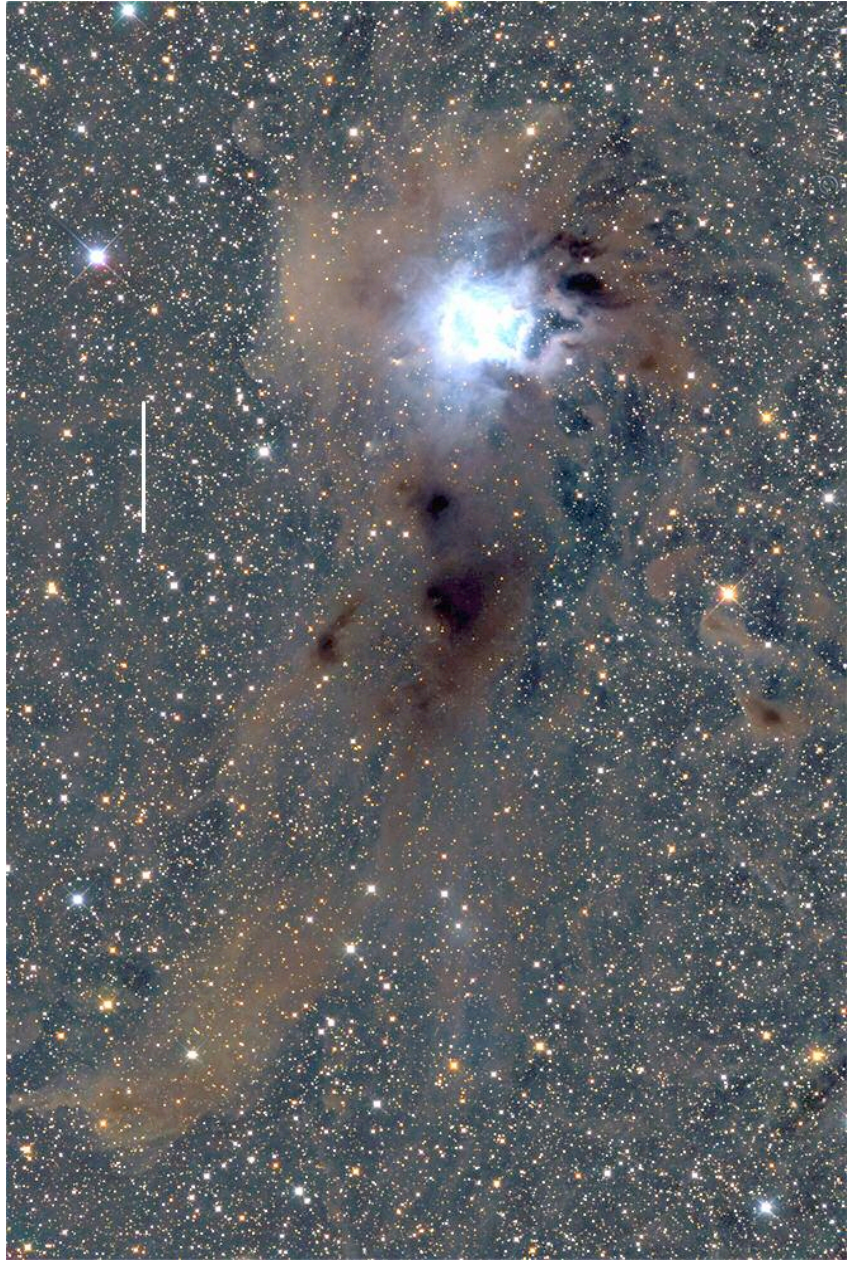

**Figure 1.** NGC 7023 in the dark clouds L1174 and L1172 in a deep optical image (tvdavisastropics.com), showing one prominent filament. The scale bar indicates 1 pc.

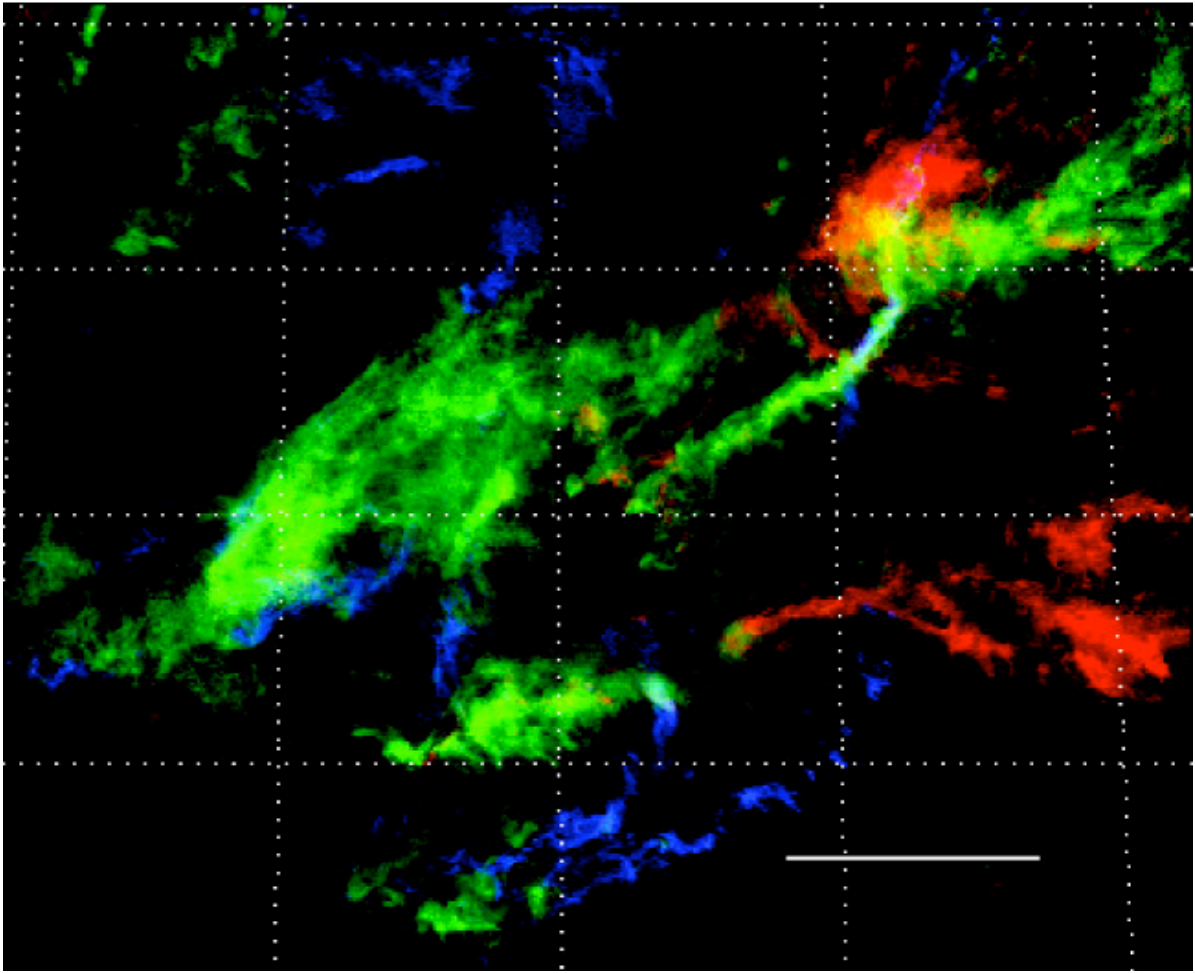

**Figure 2.** Taurus complex in $^{13}$CO J=1-0, showing the hub L1495 in the NW radiating three filaments, and three "neighbor filament" features with similar length and orientation (Goldsmith et al 2008). Velocities $V_{LSR}$=3-5 km s$^{-1}$ are in blue, 5-7 km s$^{-1}$ in green, and 7-9 km s$^{-1}$ in red. The scale bar indicates 5 pc.

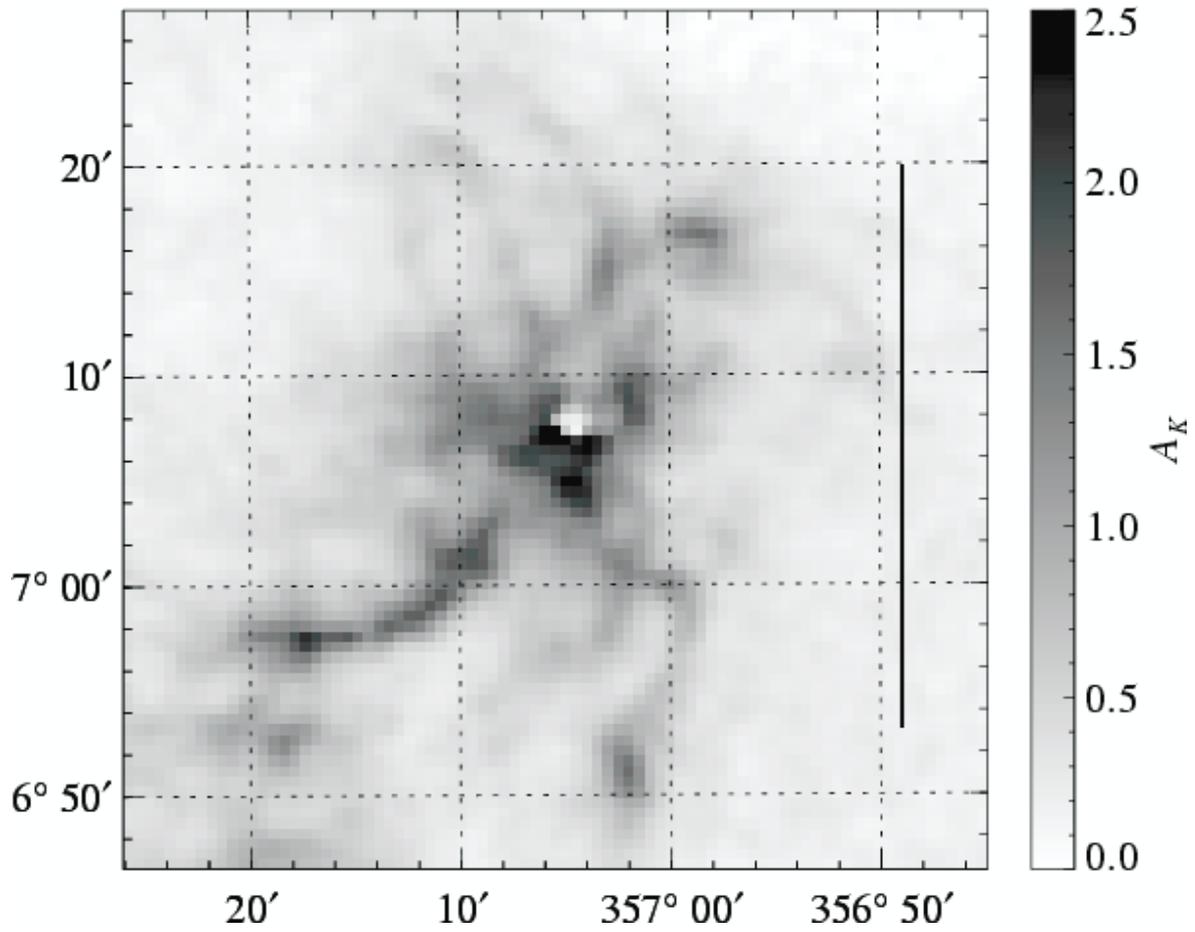

**Figure 3.** $A_K$ image of Barnard 59 in the Pipe Nebula, based on near-infrared star counts (Lombardi et al 2006), showing two curving filaments to the S and SE, and one to the NW. The central white spot is an artifact. The scale bar indicates 1 pc.

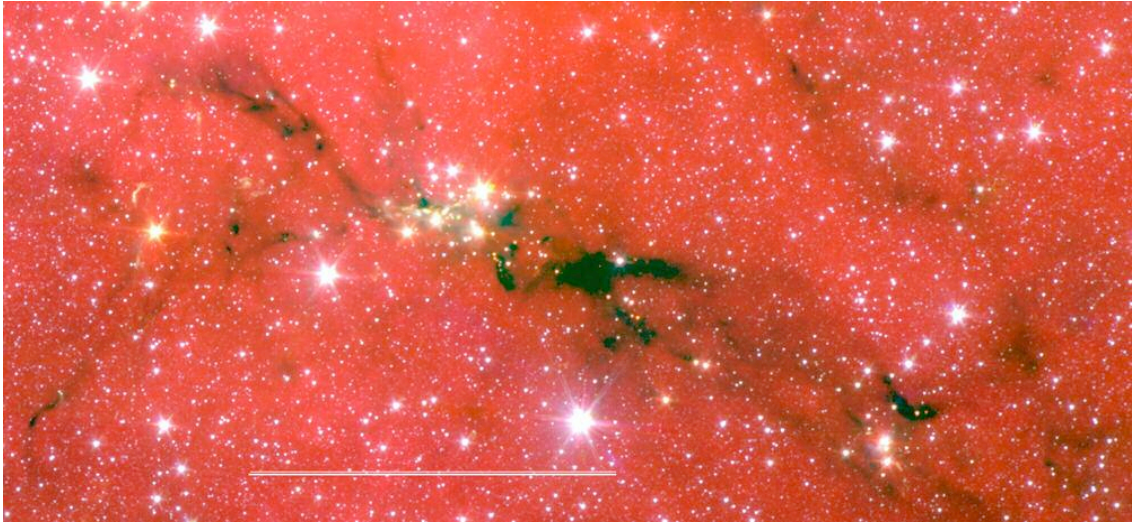

**Figure 4.** Serpens South cluster, from IRAC bands 1 and 2 (Gutermuth et al 2008c), with N to the right and E to the top. The young cluster radiates two thin filaments to the S and SE, and a broader feature to the NW. This NW feature appears to join a second hub-filament system to the N. The scale bar indicates 0.5 pc.

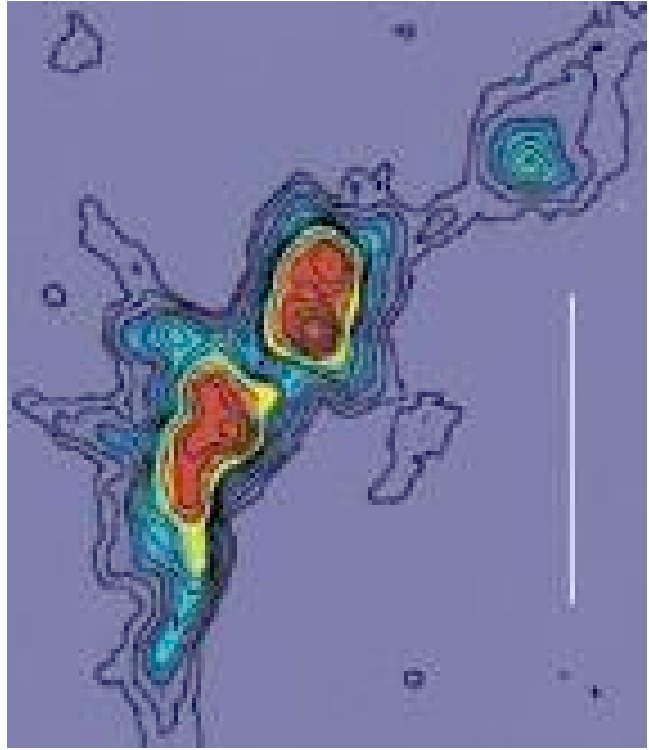

**Figure 5.** The Serpens "Cluster A" at 1.1 mm (Enoch et al 2007), showing an elongated hub and with two long and two short filaments. The scale bar indicates 0.5 pc.

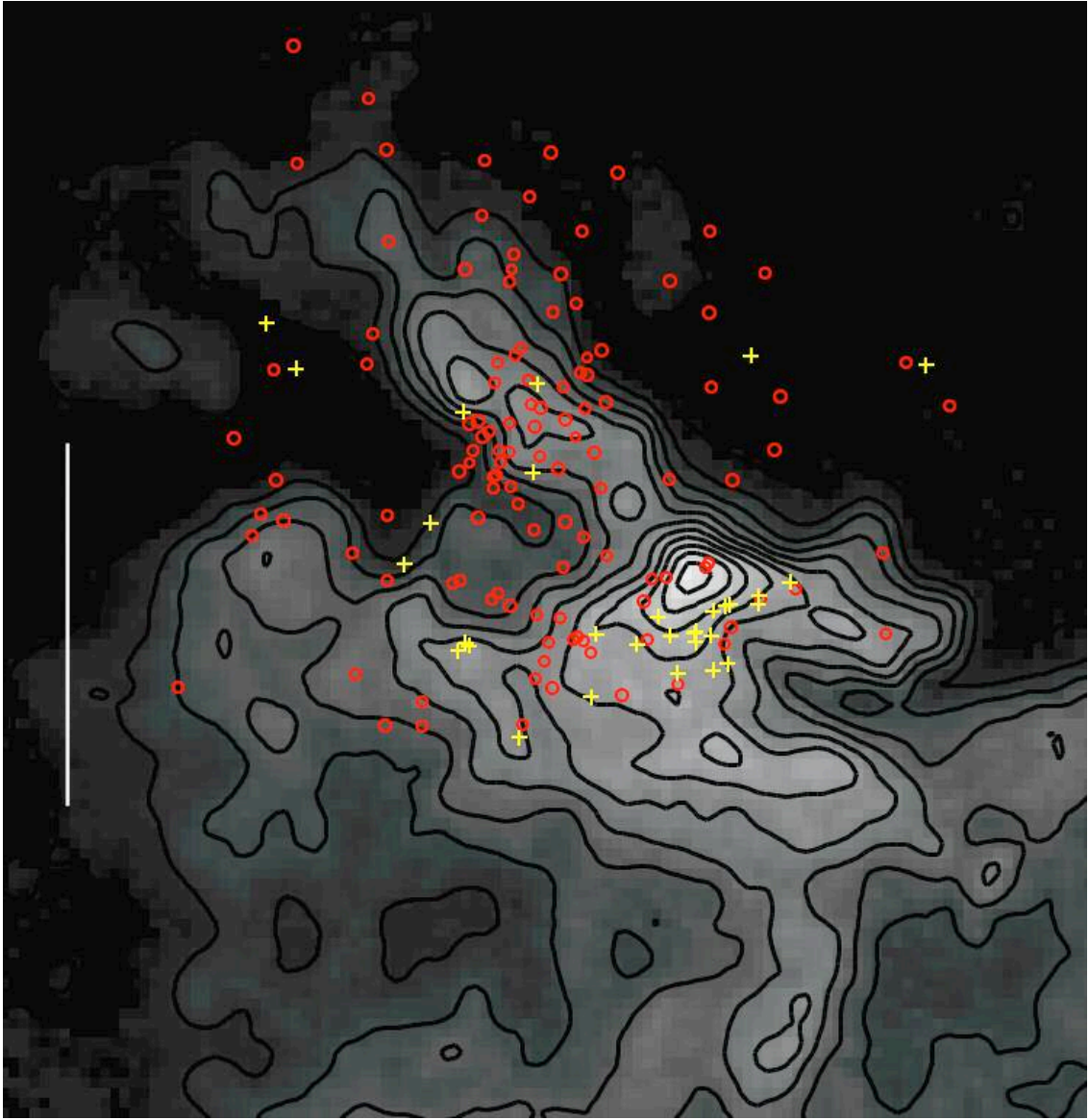

**Figure 6.** The cluster IC348 in Perseus, with Class II YSOs (red circles) and Class I protostars (yellow crosses) based on *Spitzer Space Telescope* observations, superposed on contours of $^{13}$CO 1-0 integrated intensity ranging from 1 to 15 K km s$^{-1}$ (Ridge et al 2006, Muench et al 2007). The dense gas emission peaks near the concentration of protostars IC348-SW, and extends along four filaments. The scale bar indicates 1 pc.

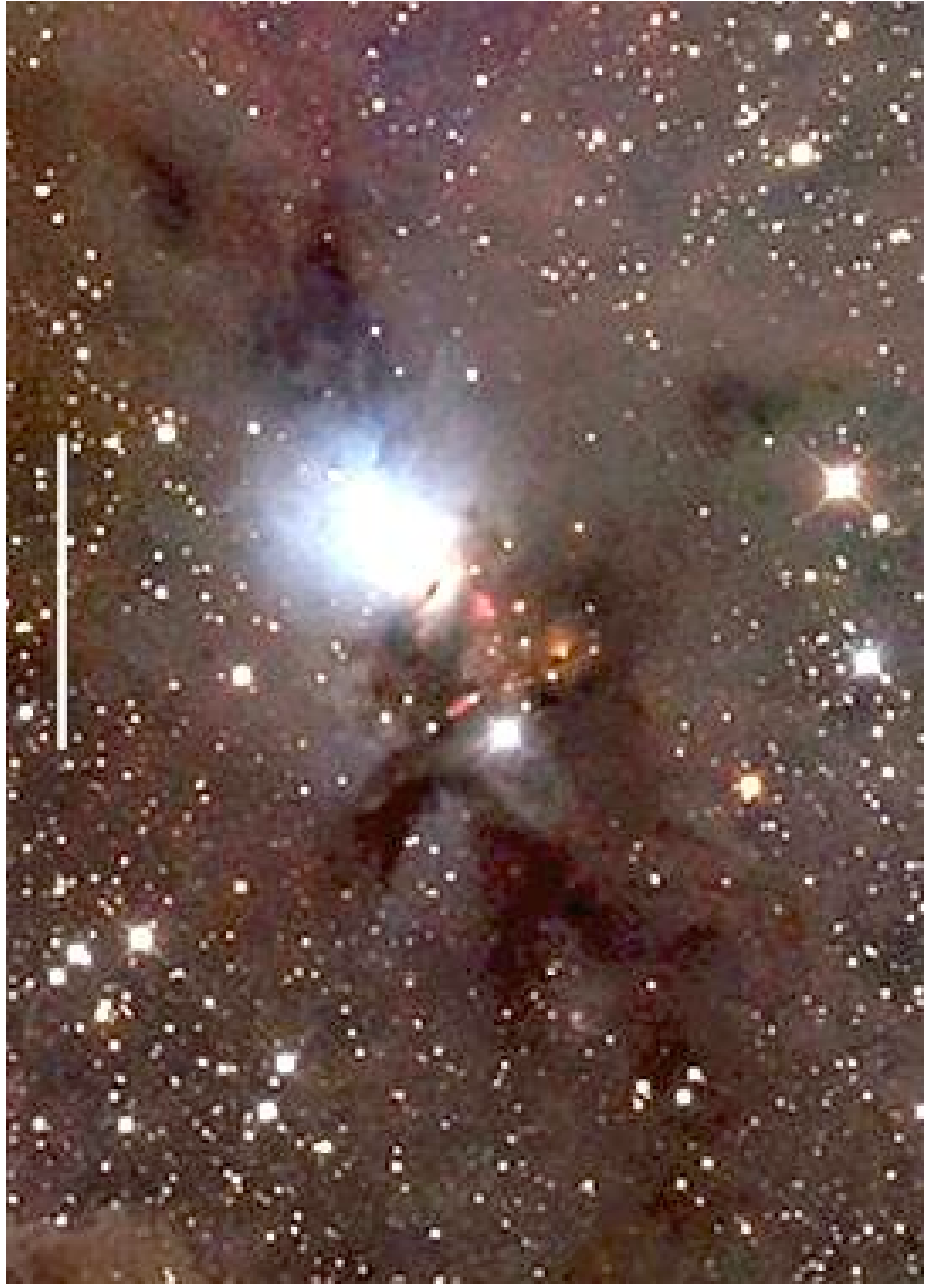

**Figure 7.** NGC 1333 in the W part of Perseus, in a deep optical image (nightskyphotography.com), showing the embedded cluster and 5 filamentary extensions. The scale bar indicates 1 pc.

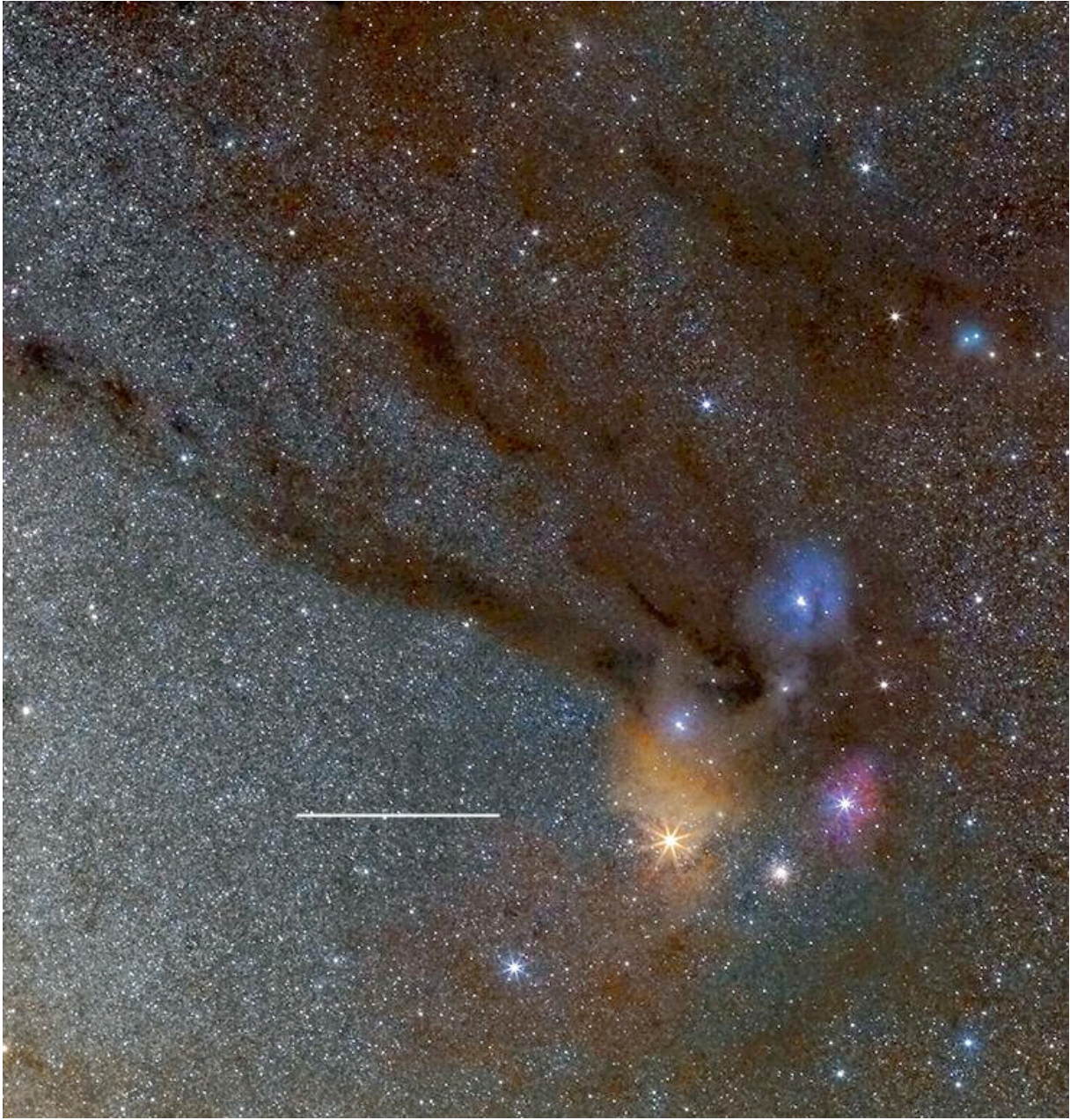

**Figure 8.**   Ophiuchus complex in a deep optical image (astromodelismo.es), showing the embedded cluster and 4 nearly parallel filaments extending to the NE, 2 curving filaments to the S, and a neighbor filament offset to the NW.  The scale bar indicates 5 pc.

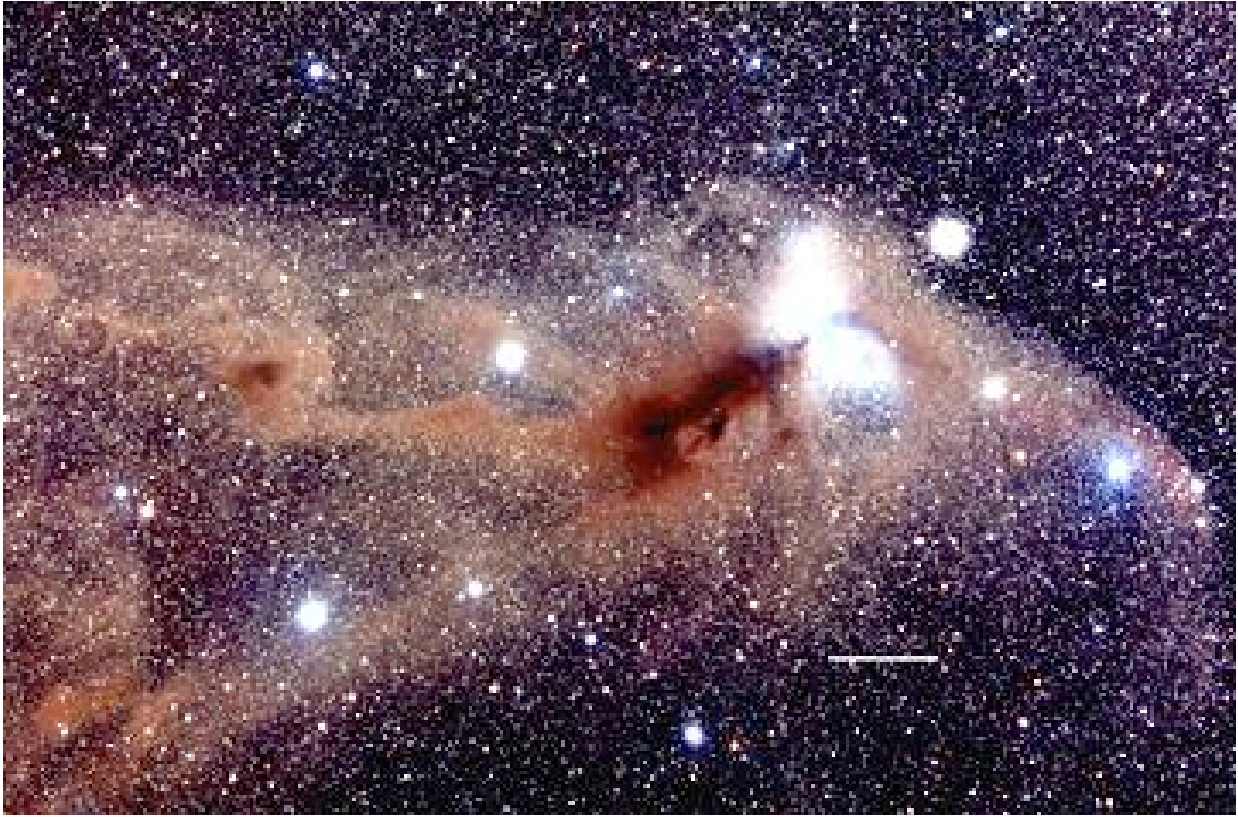

**Figure 9.** The Corona Australis cloud in a deep optical image (starryscapes.com), showing emission nebulosity from the Coronet cluster, five filaments extending to the NE, and three to the SW. The scale bar indicates 1 pc.

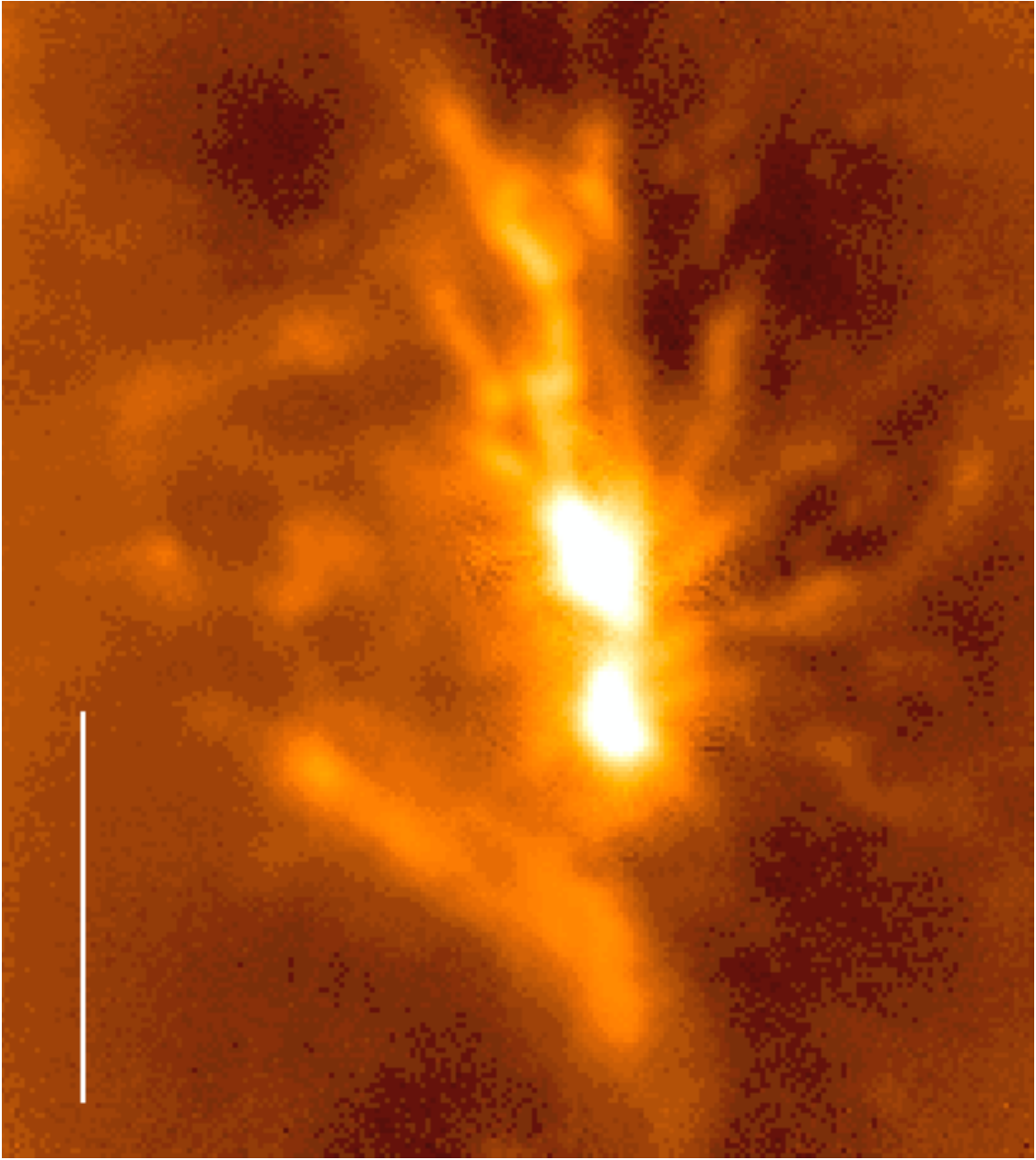

**Figure 10.** The integral filament associated with the Orion Nebula Cluster, in a submillimeter continuum image at 0.85 mm (Johnstone & Bally 1999). The bright elongated hub radiates four arms to the W and three to the N, in addition to the "bar" in the S. The scale bar indicates 0.5 pc.

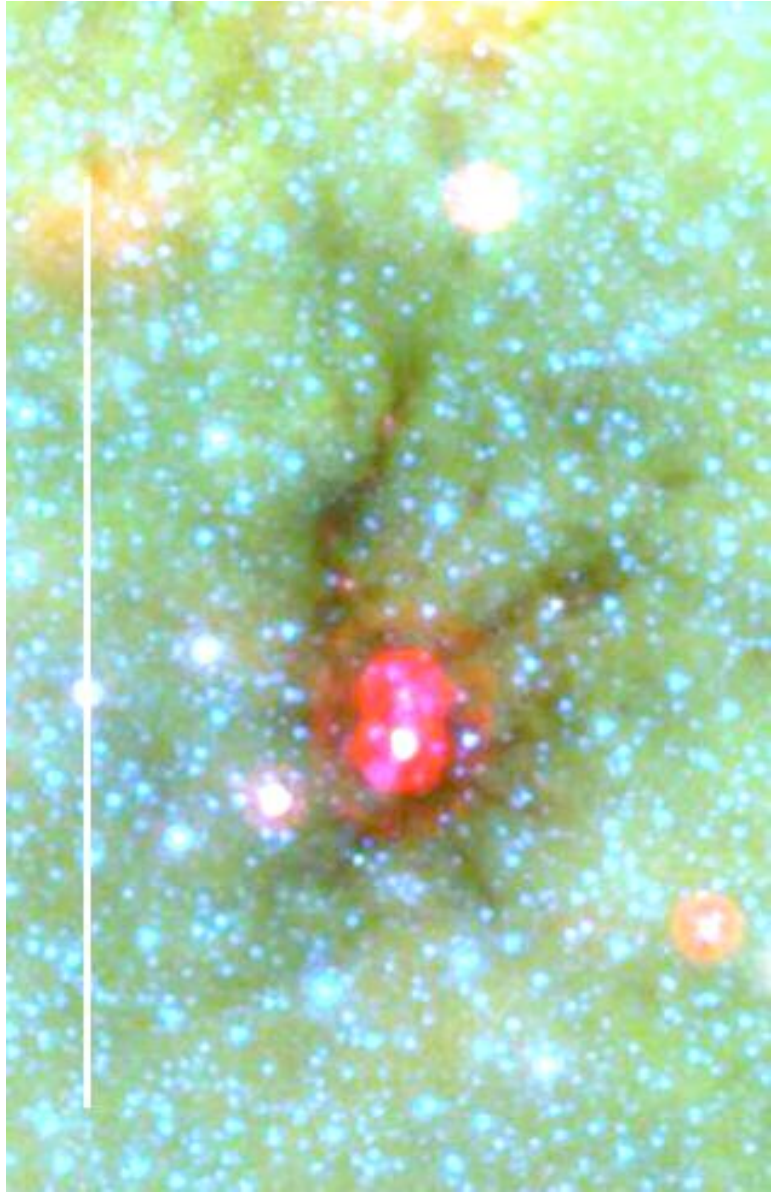

**Figure 11.** GLIMPSE/MIPSGAL image of the IRDC associated with MIPS source G335.59-0.29 (alienearths.org/glimpse), showing two candidate massive protostars and four filaments radiating to the galactic NW. Here blue, green, and red indicate respectively emission in the bands at 3, 8, and 24 $\mu$m. The scale bar indicates 5 pc.

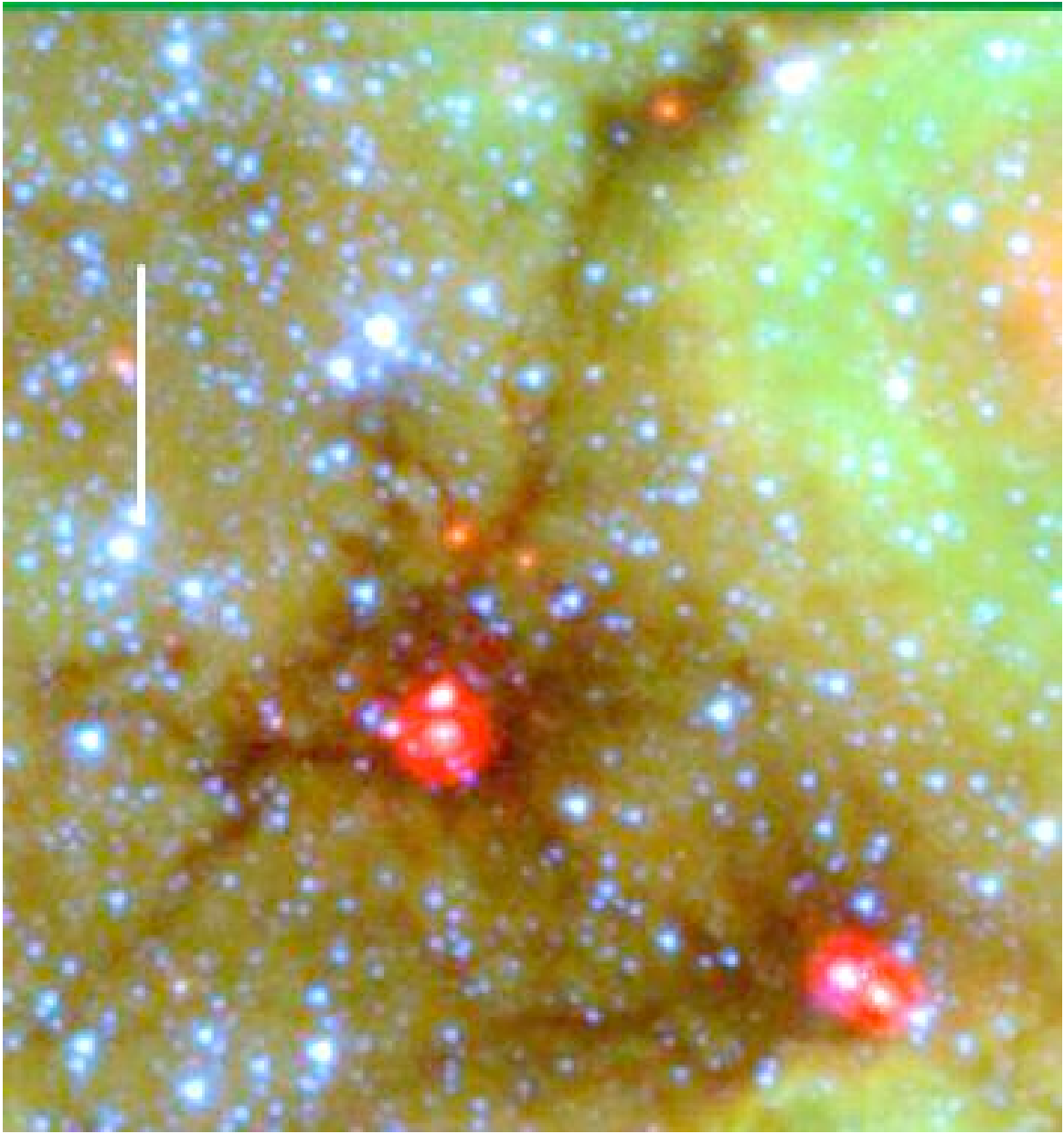

**Figure 12.** GLIMPSE/MIPSGAL image of two hub-filament systems with central massive protostars. The central object is G343.76-0.16, showing three filaments to the N and one to the S. The SE object is G343.72-0.18, showing two candidate massive protostars, three filaments radiating to the N and two to the E. The scale bar indicates 1 pc.

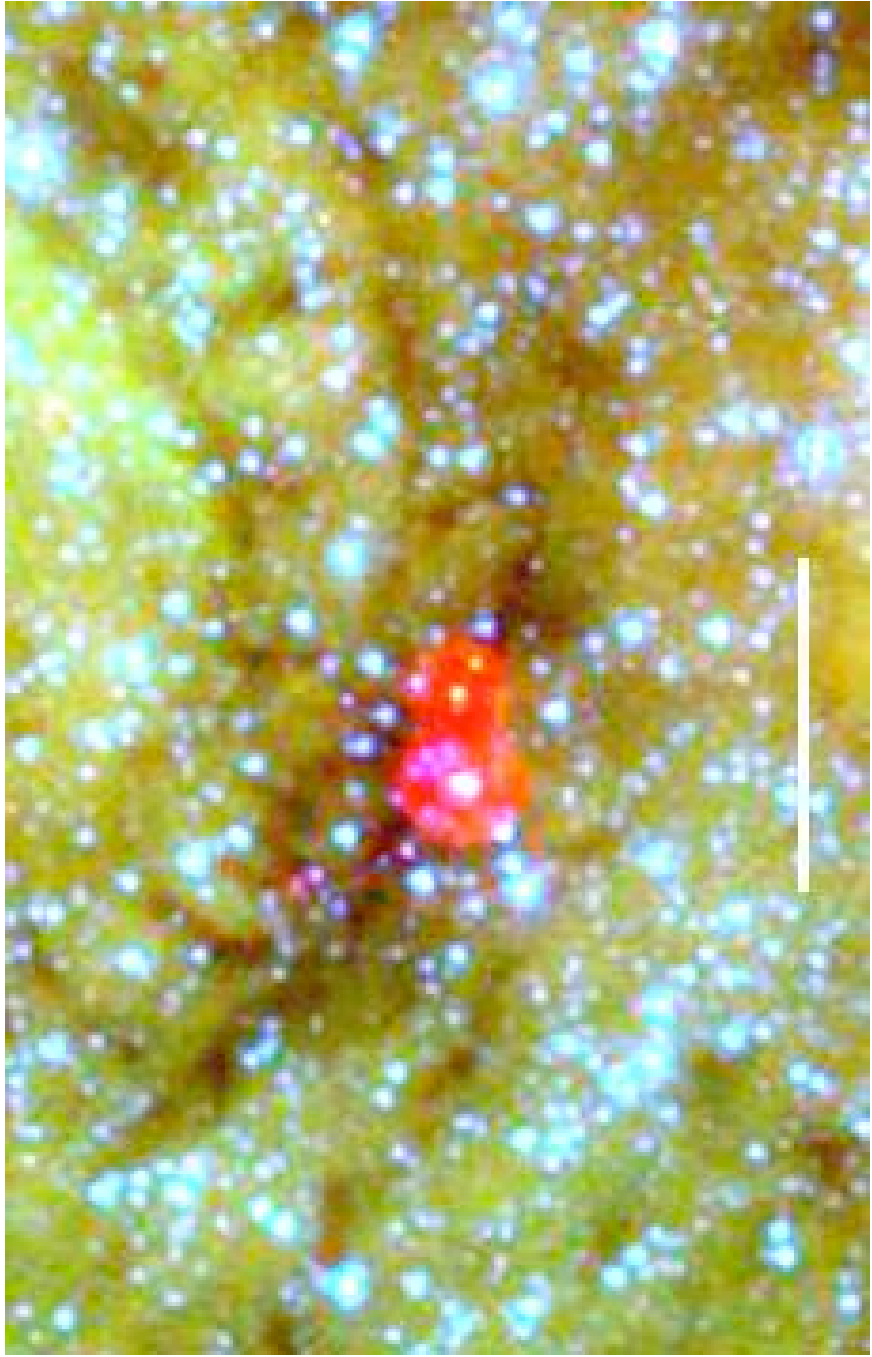

**Figure 13.** GLIMPSE/MIPSGAL image of G343.78-0.24, showing three candidate massive protostars, one filament radiating to the N, two to the E, and two to the S. Two curving neighbor filaments extend to the NE. The scale bar indicates 1 pc.

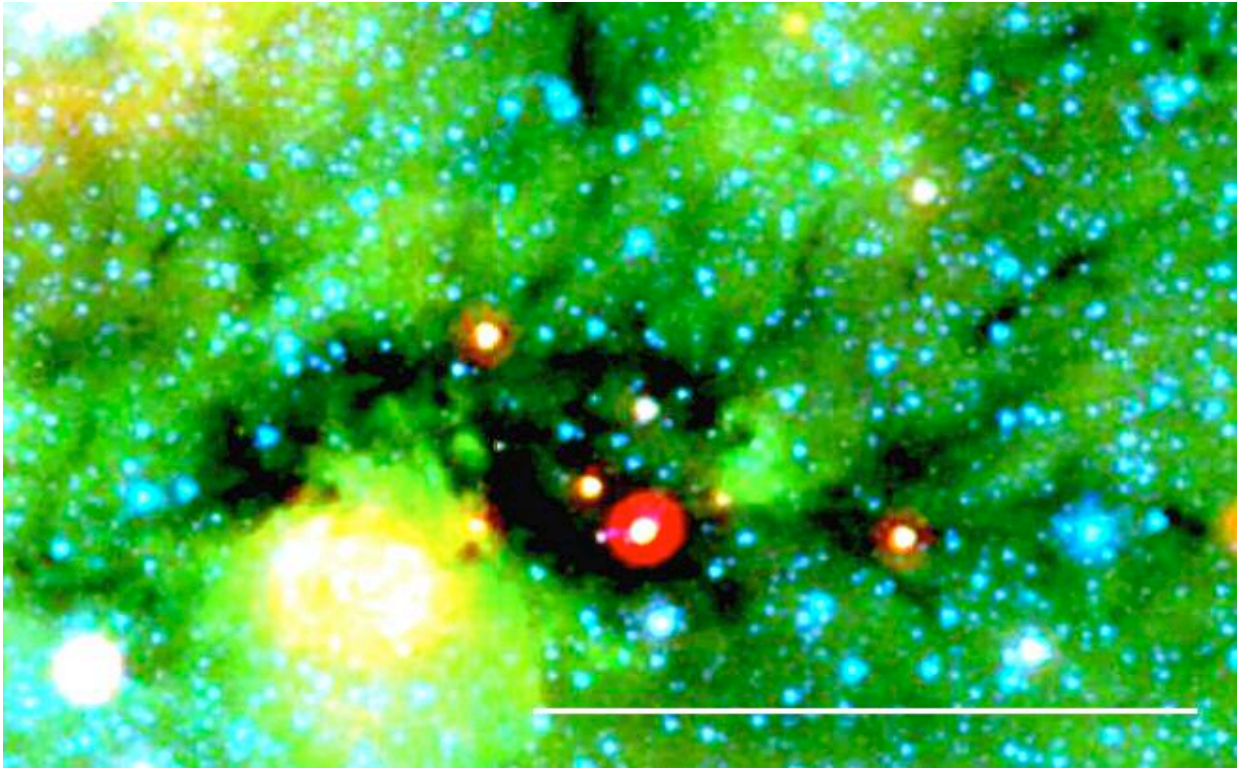

**Figure 14.** GLIMPSE/MIPSGAL image of G335.43-0.24, showing a candidate massive protostar and an elongated hub with six filaments radiating to the NW, one to the W, and two to the SW. The scale bar indicates 5 pc.

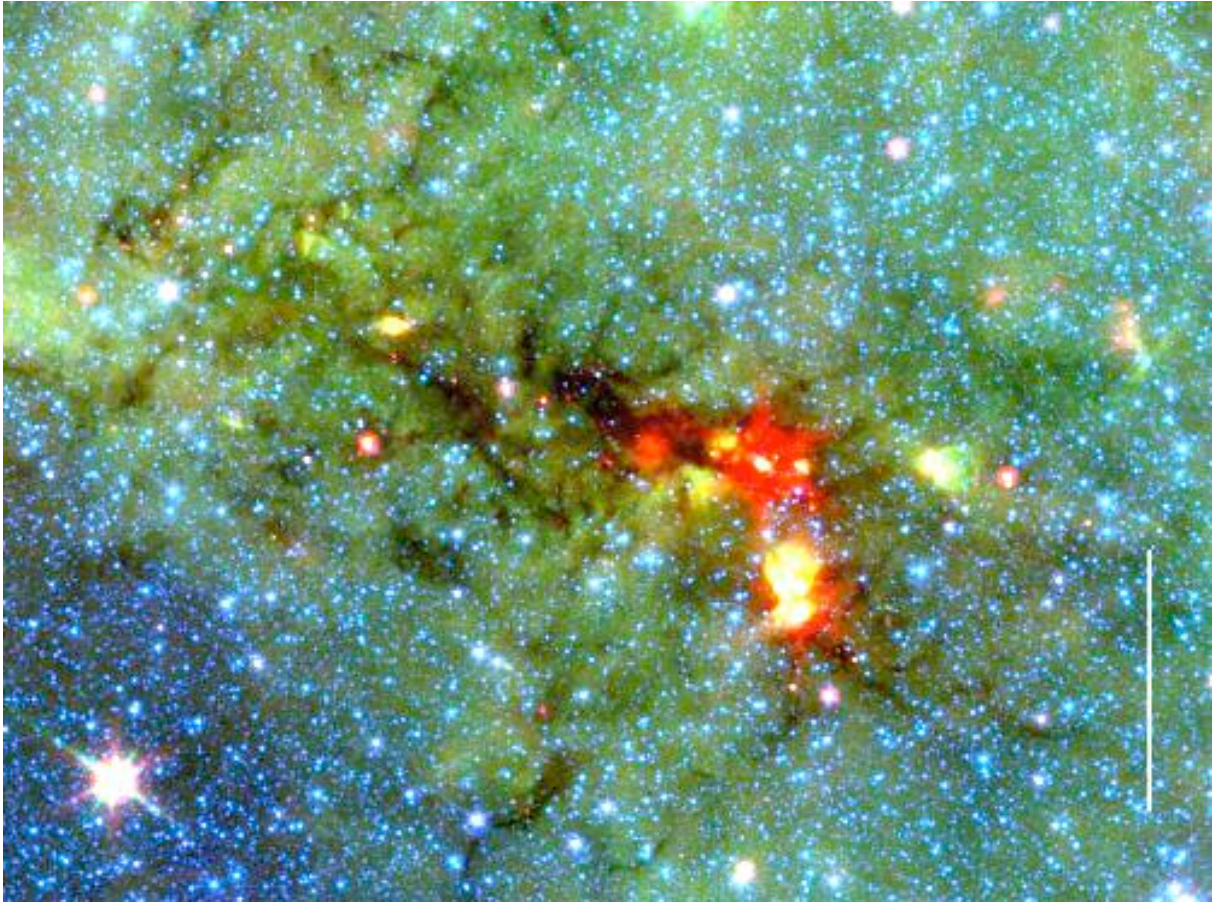

**Figure 15.** GLIMPSE/MIPSGAL image of G345.00-0.22, showing two candidate massive protostars and several fainter sources, in an elongated hub with seven distinct and two diffuse filaments radiating to the N and NE and additional diffuse filaments radiating to the S. The scale bar indicates 5 pc.

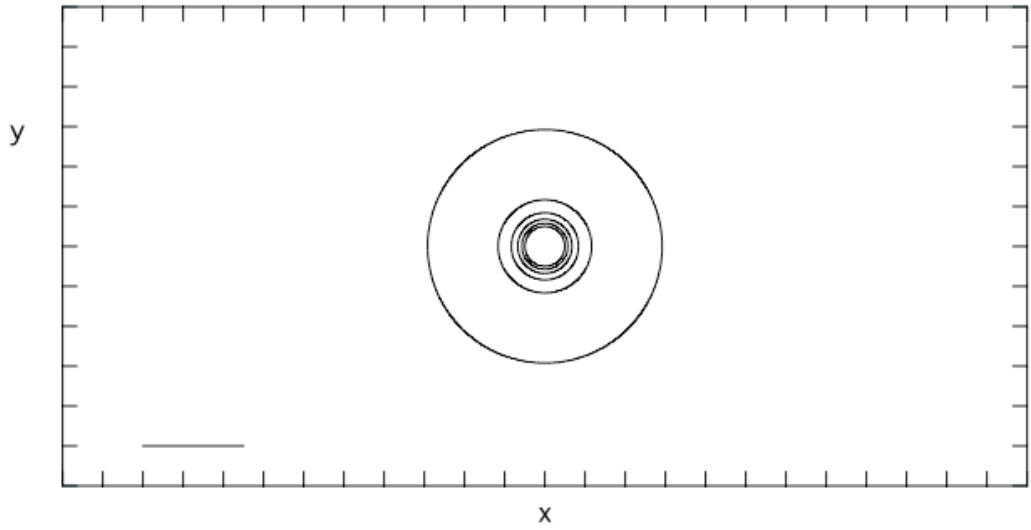

**Figure 16.** Unmodulated column density contours of a compressed clumpy medium with $N_u = 3.0 \times 10^{21}$ cm$^{-2}$, $n_u = 200$ cm$^{-3}$, $n_0 = 5000$ cm$^{-3}$, and g=1. Contours start at $3.2 \times 10^{21}$ cm$^{-2}$ and increase in steps of $1.0 \times 10^{21}$ cm$^{-2}$. Ticks have spacing 2 pc, and the scale bar indicates 5 pc.

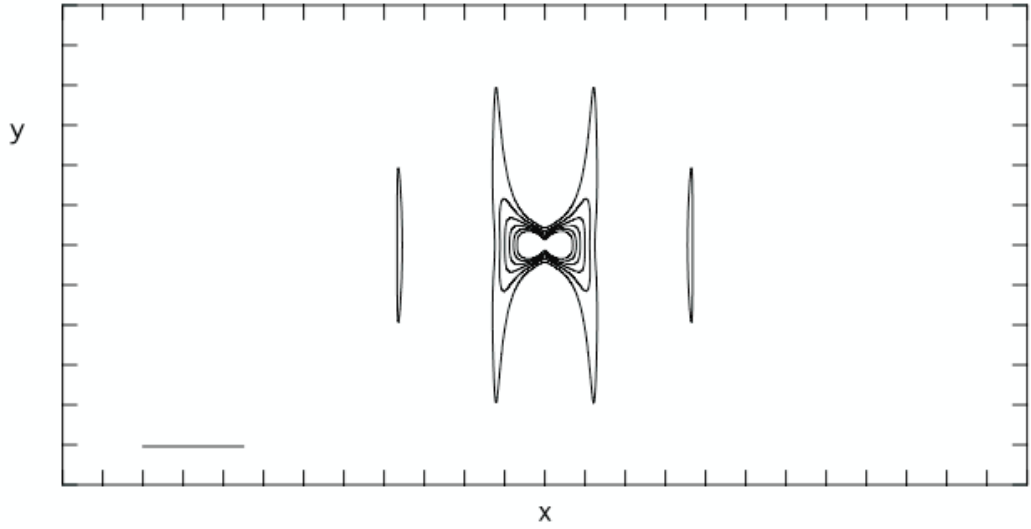

**Figure 17.** Modulated column density contours of a compressed clumpy medium with A=0.34 and other parameters as in Figure 16. The lowest contour is $5.25 \times 10^{21}$ cm$^{-2}$ and contours increase in steps of $1.00 \times 10^{21}$ cm$^{-2}$. Tick marks have spacing 2 pc, and the scale bar indicates 5 pc. Here relatively few filaments converge on a hub of low aspect ratio, as seen in Figures 2, 3, 4, 6, 11, 12, and 13. Less extended "neighbor filaments" are also seen as in Figures 2, 8, and 13.

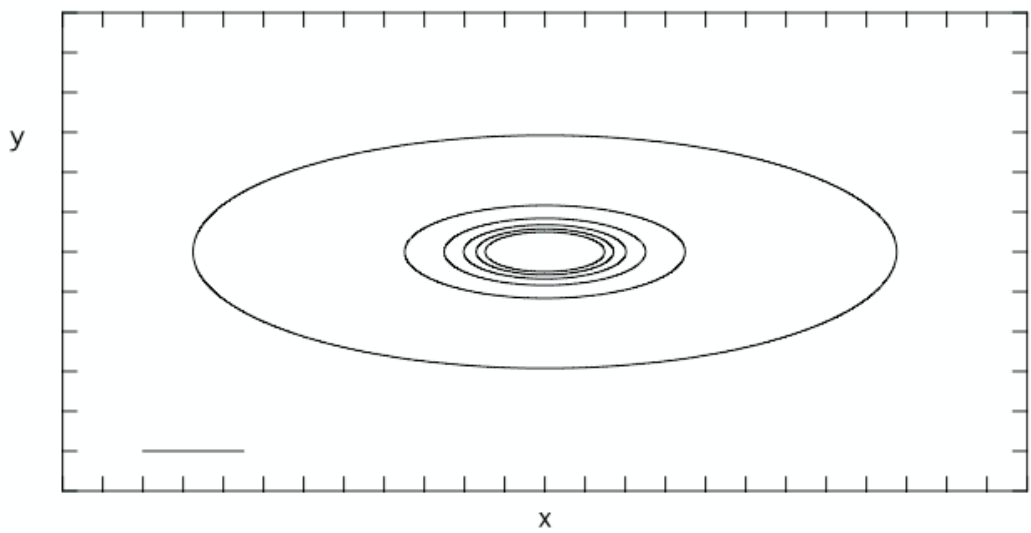

**Figure 18.** Unmodulated column density contours of a compressed clumpy medium as in Figures 16-17 but with elongation parameter g=3. As in Figure 16, contours start at $3.2 \times 10^{21}$ cm$^{-2}$ and increase in steps of $1.0 \times 10^{21}$ cm$^{-2}$. Ticks have spacing 2 pc and the scale bar indicates 5 pc.

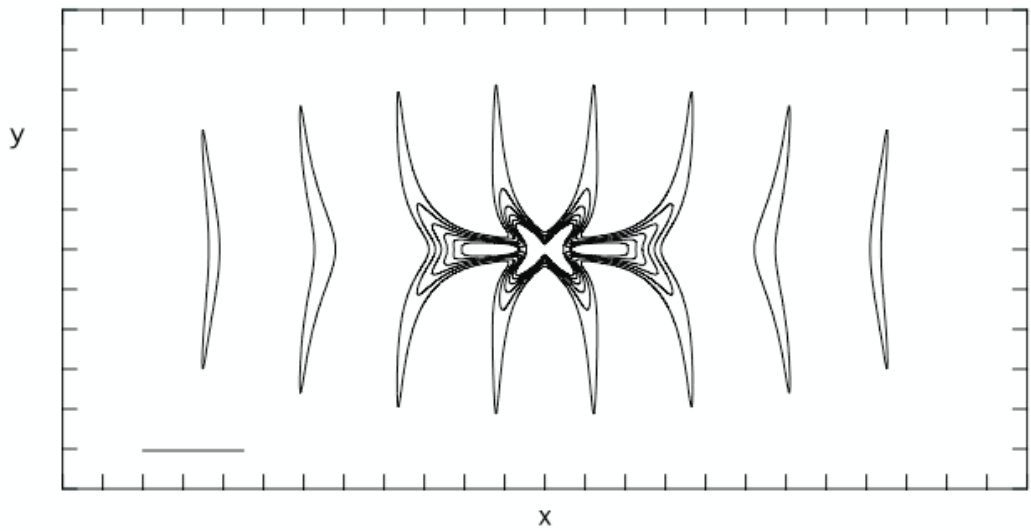

**Figure 19.** Modulated column density contours of a compressed clumpy medium with parameters as in Figures 16-17 but with elongation parameter g=3. As in Figure 17, contours start at $5.25 \times 10^{21}$ cm$^{-2}$ and increase in steps of $1.00 \times 10^{21}$ cm$^{-2}$. Tick marks have spacing 2 pc and the scale bar indicates 5 pc. Here the most prominent filaments have regular spacing and are nearly parallel. They converge on an elongated hub as seen in Figures 8, 9, 14, and 15.

**Table 1.** Hubs and filaments associated with the nearest young stellar groups

| Hub | Radiating Filaments | Neighbor Filaments | $N_{max}$ ($10^{21}$ cm$^{-2}$) | N(YSO) | R (pc) | Protostar Fraction | D (pc) | Refs |
|---|---|---|---|---|---|---|---|---|
| (1) | (2) | (3) | (4) | (5) | (6) | (7) | (8) | (9) |
| NGC7023 | 1 | 0 | 15 | 28 | 0.5 | 0.2 | 290 | 1,2 |
| L1495 | 3 | 3 | 17 | 29 | 0.3, 0.5 | 0.3 | 140 | 3,4 |
| B59 | 3 | 0 | 24 | 20 | 0.3 | 0.1 | 130 | 5,6,7 |
| Serpens S | 3 | 0 | 30 | 37 | 0.2 | 0.8 | 260 | 8,9 |
| Serpens | 4 | 0 | 21 | 55 | 0.2 | 0.4 | 260 | 10,4 |
| IC348-SW | 4 | 1 | 13 | 26 | 0.3 | 0.4 | 250 | 11,4 |
| NGC1333 | 5 | 0 | 20 | 96 | 0.4 | 0.3 | 250 | 12,4 |
| L1688 | 6 | 1 | 34 | 117 | 0.2, 0.4 | 0.3 | 125 | 13,4 |
| CrA | 8 | 0 | 21 | 24 | 0.3 | 0.3 | 170 | 14,4 |

References: 1, tvdavisastropics.com; 2, Kirk et al (2009); 3, Goldsmith et al (2008); 4, Gutermuth et al (2008a); 5, Lombardi, Alves & Lada (2006); 6, Brooke et al (2007); 7, Alves, Lombardi & Lada (2007); 8, Gutermuth et al (2008b); 9, R. Gutermuth, personal communication; 10, Enoch et al (2007); 11, Muench et al (2007); 12, nightskyphotography.com; 13, astromodelismo.es; 14, starryscapes.com

**Table 2.** Input parameters

________________________________________________________

| | |
|---|---|
| Initial medium density $n_u$ (cm$^{-3}$) | 200 |
| Initial peak density $n_0$ (cm$^{-3}$) | 5000 |
| Initial medium column density, $N_u$ (10$^{21}$ cm$^{-2}$) | 3.0 |
| Velocity dispersion $\sigma$ (km s$^{-1}$) | 0.84 |
| Modulation amplitude $A$ | 0.34 |
| Prolate spheroid aspect ratio $g$ | 1, 3 |

________________________________________________________